# Training Ecosystems: A Computational Approach to Uncovering Learning Behavior in Unconventional Contexts

**Adrita Samanta**[1], **Hananel Hazan**[2], and **Michael Levin**[2*]

[1] Greater Commonwealth Virtual School, Greenfield, MA, USA
[2] Allen Discovery Center at Tufts University, Medford, MA, USA

[*] Author for Correspondence: michael.levin@tufts.edu

**Abstract**

Recent progress in diverse intelligence has shown simple learning capacities below the organism level – single cells and even molecular networks. However, there are still many knowledge gaps around learning capacity above the organism level, and about memory implemented purely by dynamical interactions without explicit memory media. We demonstrate that minimal ecological dynamics (in silico) are sufficient for several kinds of learning, assayed as changes in both, magnitude of response, and of recovery time. Systematic exploration of over 220,000 parameter combinations in a simulated classic predator-prey model revealed that, when perturbed by stimuli, recovery time exhibits habituation, sensitization, and a form of discrete number learning in a scale-invariant manner. Robustness analysis revealed that habituation and sensitization persist under stochastic perturbations, while discrete number learning is disrupted even at low noise levels. Dimensionality reduction revealed that the incidence of learning capacity is primarily determined by ecological interaction strengths. Clear, unique clustering patterns in parameter space allow high prediction accuracy for novel parameter combinations that enable learning. Response magnitude revealed a striking asymmetry: 90.6% of parameter combinations exhibited recovery time sensitization paired with habituation of response magnitude, while the opposite pattern was extremely rare. These findings highlight a set of phenomena at the intersection of ecology, basal cognition, and mathematics with many implications for a wide range of systems describable by similar kinds of equations. These properties provide numerous efforts in biology and engineering with a substrate that has considerable, pre-patterned, propensity for learning, which ultimately arises from mathematics, not depending on the details of physics or biology.

## 1 Introduction

Biological systems need to adapt to repeated environmental disturbances to survive and stay stable. Classic adaptation patterns like habituation (decreased responsiveness) and sensitization (increased responsiveness) are well-documented across scales, from cellular signaling to neural circuits to whole organisms [1, 2]. But we still don't fully understand the mechanisms behind these adaptive responses, especially whether they can emerge from simple dynamical interactions. More broadly, the field of Diverse Intelligence seeks fundamental insight into what features are necessary and sufficient for memory, and learning can be recognized in truly unconventional substrates [3-8]. Especially challenging is the idea that learning does not require brains [9-15], and the development of tools to help recognize learning phenomena at very small or very large spatio-temporal scales, or in behavioral spaces different from the conventional 3D space of motile behavior [16].

Our perspective is that learning should be a nearly ubiquitous, not rare and neural-specific, feature in the world. Recent work shows that even simple biological systems can do complex computations and show memory-like properties [5, 17, 18], as can materials [19, 20]. To spur the development of tools and approaches for recognition of learning effects in diverse systems, we sought to analyze an in-silico model of large-scale phenomena, asking: can ecological systems, governed by simple population dynamics between two species, show recognizable forms of learning and pattern recognition through repeated perturbations? We hypothesized that recovery dynamics would change across multiple disturbance cycles in a way that would match well-known types of learning in classical behavioral science [9, 21-24]. Having shown this, we sought to ask what system parameters determine different adaptive behaviors, and how common these are in the space of possibilities. We view this analysis as providing two contributions. One is a proof of concept of how to test almost any in silico model for learning capabilities, using a behaviorist approach which focuses on functional outcomes not restricted to specific mechanisms or embodiments [25].

The second contribution concerns the specific findings that could be relevant to ecology and evolutionary biology *per se*. Learning at the population level has been studied in computational models of evolution [26-31] and in the behavioral collective intelligence of swarms [32-43]. However, these studies focus on learning through evolutionary processes or behavioral coordination, leaving open the question of whether population dynamics alone, without evolution or complex behavioral rules, can show learning and memory through purely dynamical interactions. Our model of ecosystem dynamics did not contain any evolutionary cycle because we wanted to focus on learning at the level of a population as an individual in one lifetime (a transformational process), not learning at the level of an ever-changing population through evolutionary timescales (a variational process). Thus, our approach to study learning dynamics in an extant predator-prey system occupies an unusual niche between learning at the level of an individual organism and that of populations changing through large numbers of generations.

Mathematical models of population dynamics give us a way to investigate adaptive responses to repeated perturbations. The Lotka-Volterra competition model, originally developed to explain predator-prey oscillations [44, 45], captures essential ecological interactions (growth, competition, and population coupling) with minimal mathematical complexity. While traditionally used to study equilibrium stability and coexistence, this framework can be extended to investigate how systems respond to sequences of identical perturbations, revealing whether recovery dynamics change systematically with repeated exposure. Crucially, we do not provide any additional memory medium – no mechanisms to store additional data beyond the classic predator-prey equations, and we fix all the parameters in each instance (no way to learn via connectionist approaches of tweaking weights between nodes).

We define memory as the ability of a system to respond to stimuli differently in the future, based on its history of past exposures to those stimuli (where the specifics of "differently" determine which, if any, established different kinds of learning are seen in a given case). How does one "stimulate" at the ecosystem level? We model a stimulus as the temporary introduction of some number of animals of a given species into the population. The introduction is made by the "experimenter", over and above the internal dynamics that govern changes in species population sizes. In other words, it is a pulsed external perturbation, in the same way that a visual or tactile input can be imposed on an animal to gauge response, or that a pulse of a specific chemical serves as a stimulus in training of molecular pathway models [46-48] or in endogenous and synthetic networks [49]. How does one assay the "response" at the ecosystem level? We quantified changes in population size, and in the time to recovery to pre-stimulus levels.

Here, we show that a simple two-species predator-prey model subjected to repeated perturbations in population size exhibits different adaptive response patterns, including classic habituation and sensitization. We also find a novel form of discrete number learning where systems transition to new recovery baselines only after experiencing a precise count of perturbations. We identify an interesting and specific pattern in the parameter space in which these phenomena occur, finding that they are not ubiquitous but also not very rare. These findings show that ecological models can exhibit interesting proto-cognitive and computational properties, including numerical pattern recognition, that arise purely from dynamical interactions without requiring neural circuitry or explicit memory mechanisms. Our data are not specific to ecological systems, as the mathematics we analyze can be used in a wide range of other contexts.

## 2 Methods

### *2.1 Model System*

We studied adaptive responses in a periodically perturbed predator–prey system using a two-species Lotka–Volterra competition model. The prey population, x(t), and predator population, y(t), evolve according to the coupled differential equations:

$$dx/dt = r_x \cdot x \cdot (1 - (x + \alpha_{xY} \cdot y) / K_x)$$
$$dy/dt = r_Y \cdot y \cdot (1 - (y + \alpha_{Yx} \cdot x) / K_Y)$$

Here, $r_x$ and $r_Y$ denote the intrinsic growth rates of the prey and predator populations, respectively. The parameters $\alpha_{xY}$ and $\alpha_{Yx}$ represent interspecific competition coefficients, capturing the inhibitory effect of one species on the growth of the other. Rather than fixing carrying capacities independently, we defined them self-consistently from the initial conditions and interaction strengths as:

$$K_x = x_0 + \alpha_{xY} \cdot y_0$$
$$K_Y = y_0 + \alpha_{Yx} \cdot x_0$$

This formulation ensures that the unperturbed system begins near equilibrium while allowing perturbations to probe the system's intrinsic recovery dynamics. All simulations were initialized with $x_0 = 25$ and $y_0 = 5$.

### *2.2 Numerical Integration*

The system of ordinary differential equations was integrated numerically using a fourth-order Runge–Kutta method with a fixed time step of $\Delta t = 0.1$. Each simulation was run for 8000 time steps, corresponding to a total simulated duration of 800 arbitrary time units, which was sufficient to capture long-term responses to repeated perturbations.

### *2.3 Periodic Perturbation Protocol*

To look for evidence of learning, the prey population was subjected to repeated pulse perturbations applied at regular intervals. Each pulse consisted of a transient increase in prey abundance, followed by a return to baseline after a short, fixed duration. Perturbations were applied periodically, with pulse magnitude and inter-pulse interval controlled by two parameters: pulse size (PS) and pulse frequency (PF). Perturbations were initiated only after an initial equilibration period, ensuring that observed responses reflected adaptation to repeated stimulation rather than transient effects associated with initialization. As per standard psychophysics and behavioral training protocols, we chose stimulus strengths that are clearly above threshold but otherwise minimal.

### *2.4 Recovery Time Measurement*

System response to each perturbation was quantified using a recovery time metric. For each pulse, recovery time was defined as the elapsed time required for the prey population to return to within 5% of its pre-perturbation baseline following pulse removal. Recovery times were measured sequentially across perturbations, yielding a recovery-time trajectory that characterizes how the system's responsiveness evolves with repeated stimulation.

### *2.5 Parameter Space Exploration*

To characterize the dependence of adaptive behavior on system parameters, we performed a comprehensive parameter sweep over six independent variables governing growth dynamics, interaction strength, and stimulation properties:

- Pulse size (PS): 5.0 to 25.0 in increments of 1.0
- Pulse frequency (PF): 350 to 700 in increments of 50
- Prey growth rate ($r_x$): 0.60 to 1.50 in increments of 0.10
- Predator growth rate ($r_\gamma$): 0.35 to 0.65 in increments of 0.05
- Competition coefficient $\alpha_{x\gamma}$: 0.80 to 1.30 in increments of 0.10
- Competition coefficient $\alpha_{\gamma x}$: 0.80 to 1.30 in increments of 0.10

These ranges were chosen to span a broad set of biologically and dynamically plausible regimes while maintaining numerical stability. In total, this grid search yielded approximately 220,000 unique parameter combinations. Each parameter combination was simulated once, producing a corresponding recovery-time trajectory for downstream analysis.

### *2.6 Classification of Adaptive Responses*

Recovery-time trajectories were classified based on their temporal structure using both trend-based and transition-based criteria.

Trend-Based Classification: To distinguish habituation, sensitization, and stable responses, we first computed a linear regression of recovery time versus stimulation number for each trajectory. Each stimulation pulse was numbered sequentially (1st pulse, 2nd pulse, 3rd pulse, etc.), and the recovery time measured after each pulse was plotted against its corresponding stimulation number, creating a trajectory across repeated exposures. The slope of this regression quantified the overall trend in recovery dynamics:

- Habituation was classified when the slope was less than −0.01, indicating a monotonic decrease in recovery time across perturbations
- Sensitization was classified when the slope was greater than +0.01, indicating a monotonic increase in recovery time across perturbations
- Stable behavior was classified when the absolute value of the slope was less than 0.01, indicating approximately constant recovery times across perturbations

Number Learning Detection: In addition to these trend-based categories, we identified number learning when recovery times exhibited discrete, step-like transitions at specific stimulation numbers. Transitions were detected using a threshold-based algorithm with the following criteria:

1. A candidate transition was identified when the percent change in recovery time between consecutive stimulations exceeded 3%
2. The transition was validated only if recovery times remained stable (within 1.5% variation) both before and after the transition for at least 2 consecutive stimulations
3. Only clean, sustained changes that met both stability criteria were classified as learning events

This dual classification approach allowed us to distinguish gradual monotonic trends (habituation/sensitization) from abrupt discrete transitions (number learning) while filtering out transient fluctuations or noisy dynamics.

### *2.7 Statistical Analysis of Number Learning Distribution*

To characterize the distribution of number learning transitions, we computed the percentage of parameter combinations exhibiting learning at each transition point (1→2 through 17→18). We fitted an exponential decay model of the form y = ae^(-bx) + c to the observed transition frequencies to establish the expected baseline pattern of learning difficulty. The model fitting was performed using scipy.optimize.curve_fit in Python [50]. Visual inspection of the data revealed three transitions (4→5, 5→6, and 6→7) that appeared inconsistent with monotonic exponential decay; these points were excluded from the curve fitting procedure to obtain a robust baseline estimate. The resulting model parameters were a = 6.27, b = 1.01, and c = 0.12. To quantify deviations from the expected pattern, we calculated the percentage elevation of observed frequencies relative to model predictions for the excluded transitions.

### *2.8 Magnitude Response Analysis*

To characterize a complementary metric of system adaptation, we measured response magnitude as the absolute change in equilibrium prey population before and after each perturbation. For each pulse, the pre-perturbation baseline was recorded immediately before pulse application, and the post-perturbation baseline was measured after recovery (when prey population returned to within 5% of pre-perturbation level). Response magnitude was defined as |post-baseline − pre-baseline|, yielding a magnitude trajectory across perturbations analogous to the recovery time trajectory.

Magnitude trajectories were classified using linear regression slope with a threshold of ±0.005 (half the threshold used for recovery time), chosen to capture weak sensitization signals while maintaining classification robustness. Magnitude habituation was classified when slope < −0.005, magnitude sensitization when slope > +0.005, and stable when |slope| ≤ 0.005. The lower threshold reflects the increased precision of equilibrium measurements compared to temporal recovery dynamics. UMAP visualization of magnitude response patterns used identical parameters (n_neighbors = 30, min_dist = 0.1) as recovery time analysis to enable direct spatial comparison.

To quantify the relationship between recovery time and magnitude response metrics, we identified all parameter combinations classified as habituation or sensitization in both analyses (n=35,140 overlapping cases from the 22,176 tested combinations). For these overlapping parameters, we constructed a 2×2 contingency matrix categorizing each combination by its recovery time response (habituation or sensitization) and magnitude response (habituation or sensitization), revealing the frequency of all four possible coupling patterns.

### *2.9 Dimensionality Reduction and Visualization*

To visualize structure in the high-dimensional parameter space, we applied Uniform Manifold Approximation and Projection (UMAP) [51] using the umap-learn Python package to the six-dimensional parameter vectors (pulse size, pulse frequency, prey growth rate, predator growth rate, and both interspecific competition coefficients). Prior to embedding, all parameters were standardized to zero mean and unit variance using StandardScaler from scikit-learn [52]. UMAP was performed with n_neighbors = 30 and min_dist = 0.1, using Euclidean distance as the metric, yielding two-dimensional embeddings used to visualize the organization of adaptive response regimes.

To validate the spatial separation of response categories in the embedded space, we generated 1,000 test parameter combinations sampled from regions of parameter space not included in the original grid search, including intermediate values (half-step and quarter-step between grid points), boundary regions, and limited extrapolation beyond the training bounds. These test points were projected into the UMAP embedding using the trained manifold transformation.

Individual spatial clusters within each response category were identified using DBSCAN (Density-Based Spatial Clustering of Applications with Noise) [53] from scikit-learn with eps = 0.5 and min_samples = 20. Convex boundaries with 10% radial buffer were computed for each identified cluster, and test point containment was assessed to quantify the robustness of parameter space partitioning.

### *2.10 Representative Example Selection*

For illustrative figures, representative parameter sets were selected from simulations exhibiting clean and well-defined behaviors. For number learning, only cases with single, isolated transitions were considered. For non-learning categories (habituation, sensitization, and stable), examples were selected to display clear monotonic trends or stable recovery baselines. Representative cases were chosen by random sampling within each category to avoid bias toward specific parameter values.

## 3 Results

### *3.1 The predator-prey system exhibits three distinct adaptive response patterns to repeated perturbations*

To characterize how ecological systems adapt to repeated environmental disturbances, we subjected a two-species Lotka-Volterra predator-prey model to periodic pulse perturbations of the prey population and measured recovery times across successive perturbations. We observed three qualitatively distinct adaptive response patterns across parameter space: habituation (recovery times decrease with repeated perturbations), sensitization (recovery times increase with repeated perturbations), and stable responses (recovery times remain constant) ( → Fig. 1). Classification was based on linear regression slopes of recovery time versus stimulation number, with thresholds set at ±0.01 to distinguish meaningful trends from noise (a slope of ±0.01 corresponds to a 10% change in recovery time across 10 perturbations for systems with baseline recovery times around 10 time units). Across approximately 220,000 parameter combinations tested, we found that 65,944 (30.0%) exhibited sensitization, 7,021 (3.2%) exhibited habituation, and 147,035 (66.8%) showed stable responses with no significant trend. Systems exhibiting habituation showed recovery times that decreased by an average of 0.08 time units per stimulation (representing 15-25% total reduction across typical perturbation sequences), while sensitization cases increased by an average of 0.12 time units per stimulation (representing 20-40% total increase). These patterns demonstrate that the same ecological system can exhibit fundamentally different forms of memory and adaptation depending solely on intrinsic system parameters, with no changes to external forcing patterns.

### *3.2 Number learning emerges as discrete transitions at specific stimulation counts*

To investigate whether ecological systems could exhibit count-based learning, we analyzed recovery time trajectories for abrupt, discrete transitions at specific stimulation numbers. We identified number learning events when recovery times exhibited step-like changes (>3% between consecutive stimulations) that were preceded and followed by stable baselines (within 1.5% variation for at least 2 stimulations). We observed clean number learning transitions at stimulation numbers ranging from 2 to 18, with systems displaying single isolated transitions at specific counts ( → Fig. 2). Of the 220,000 parameter combinations tested, 6,821 (3.1%) exhibited at least one discrete number learning transition, with 6,234 systems (91.4%) showing exactly one transition, 512 systems (7.5%) showing two transitions, and 75 systems (1.1%) showing three or more transitions. Number learning occurred independently of trend-based classifications: learning systems could exhibit habituation, sensitization, or stable baselines both before and after the discrete transition. These results demonstrate that ecological systems can spontaneously develop numerical pattern recognition, transitioning to new stable recovery baselines only after experiencing a precise number of perturbations.

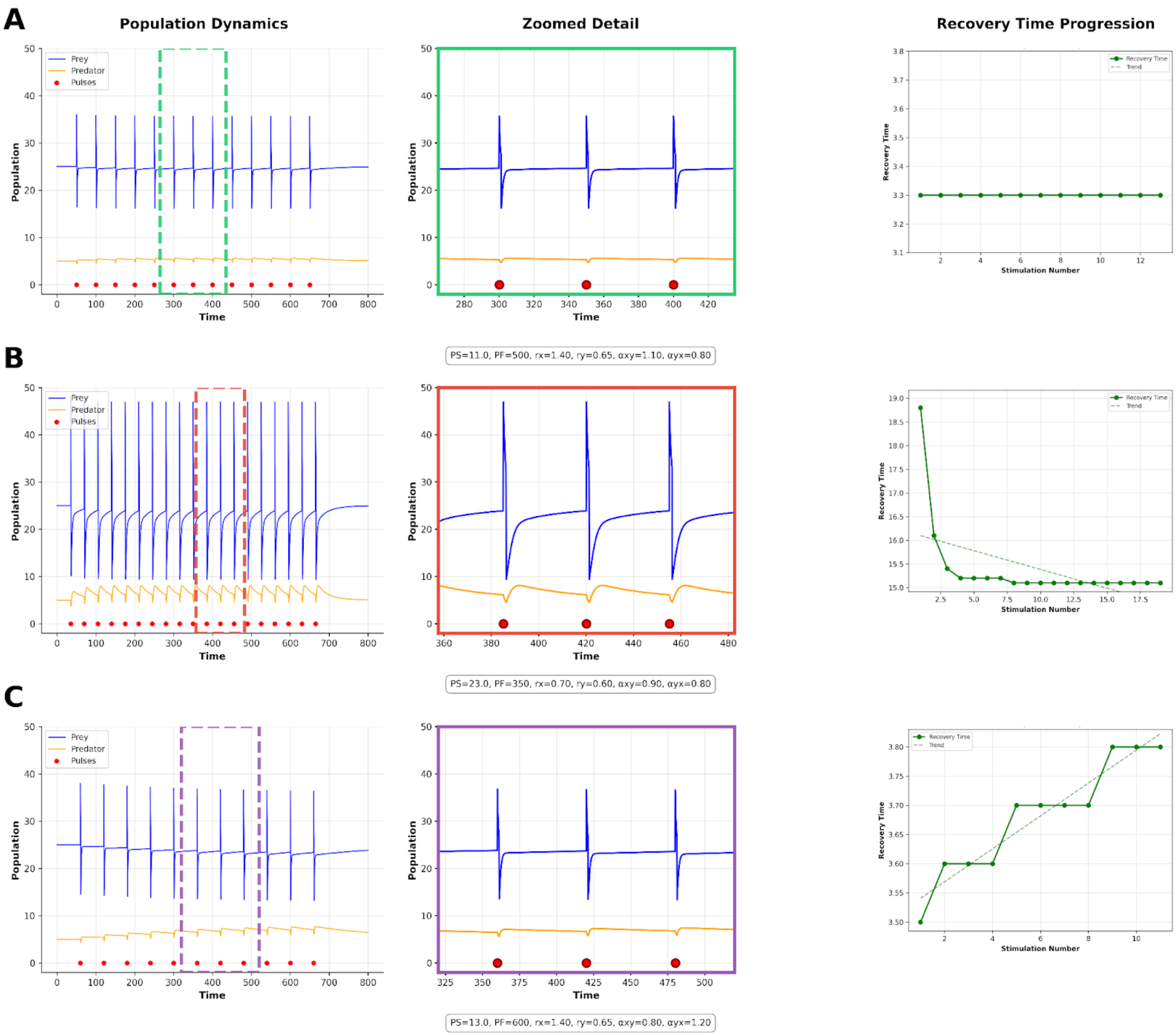

**Fig. 1** Overall adaptive response categories in a periodically perturbed predator-prey system. Each row shows three views of system behavior: population dynamics (left), zoomed detail of selected perturbations (middle), and recovery time progression across stimulations with trend line (right). The zoomed regions (green, red, purple boxes) in the population dynamics correspond to the detailed views in the middle column. (A) Stable: Recovery times remain constant across perturbations (PS=11.0, PF=500, $r_x$=1.40, $r_y$=0.65, $\alpha_{xy}$=1.10, $\alpha_{yx}$=0.80). (B) Habituation: Recovery times decrease with repeated perturbations (PS=23.0, PF=350, $r_x$=0.70, $r_y$=0.60, $\alpha_{xy}$=0.90, $\alpha_{yx}$=0.80). (C) Sensitization: Recovery times increase with repeated perturbations (PS=13.0, PF=600, $r_x$=1.40, $r_y$=0.65, $\alpha_{xy}$=0.80, $\alpha_{yx}$=1.20).

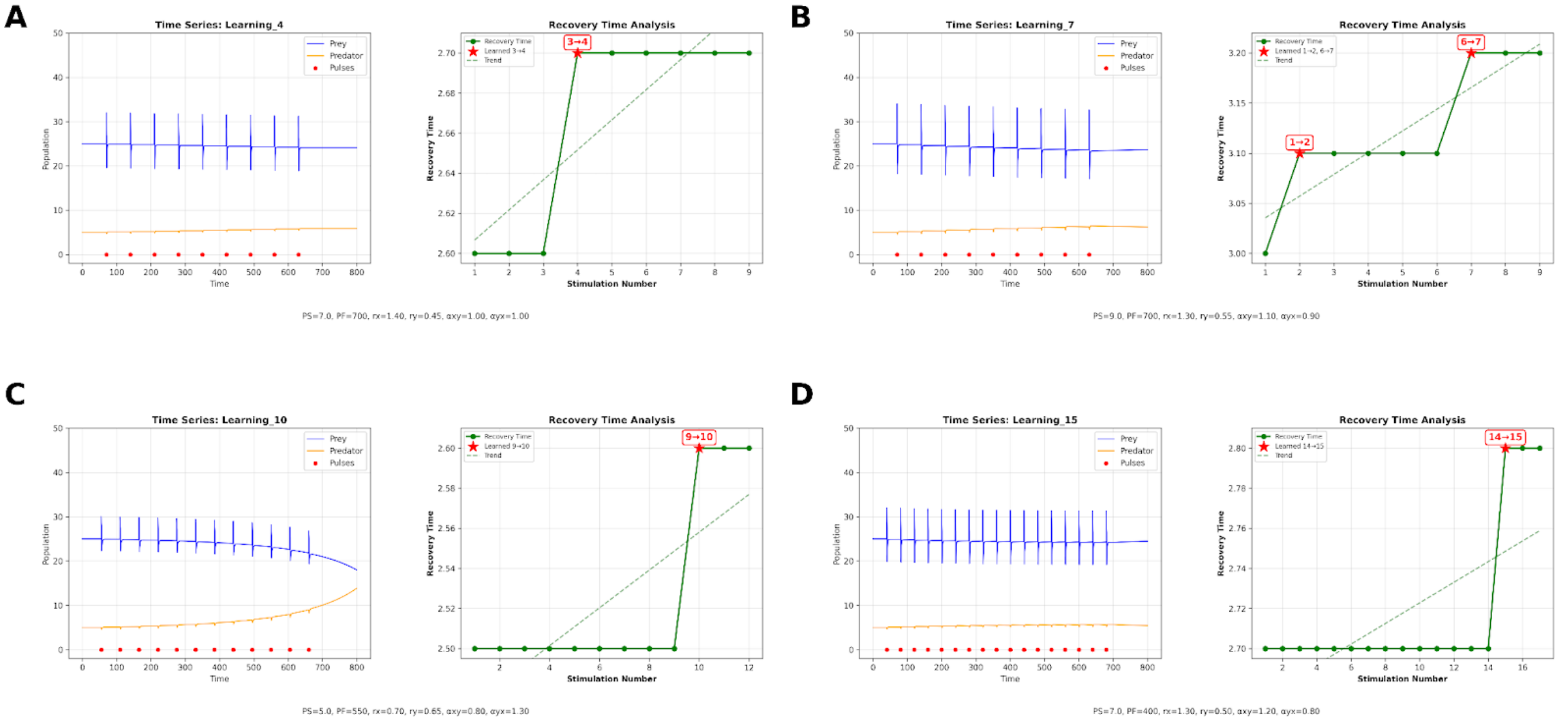

**Fig. 2** Discrete number learning transitions at specific stimulation numbers. Left: Population time series showing prey (blue), predator (orange), and pulse perturbations (red dots). Right: Recovery time analysis with transition points marked by red stars and labeled boxes. (A) Learning at stimulation 4 (3→4 transition). (B) Multiple learning events at stimulations 2 and 7 (1→2, 6→7 transitions). (C) Learning at stimulation 10 (9→10 transition). (D) Learning at stimulation 15 (14→15 transition). The system exhibits abrupt transitions to new stable recovery baselines at diverse stimulation numbers (observed range: 2-18), demonstrating tunable numerical pattern recognition.

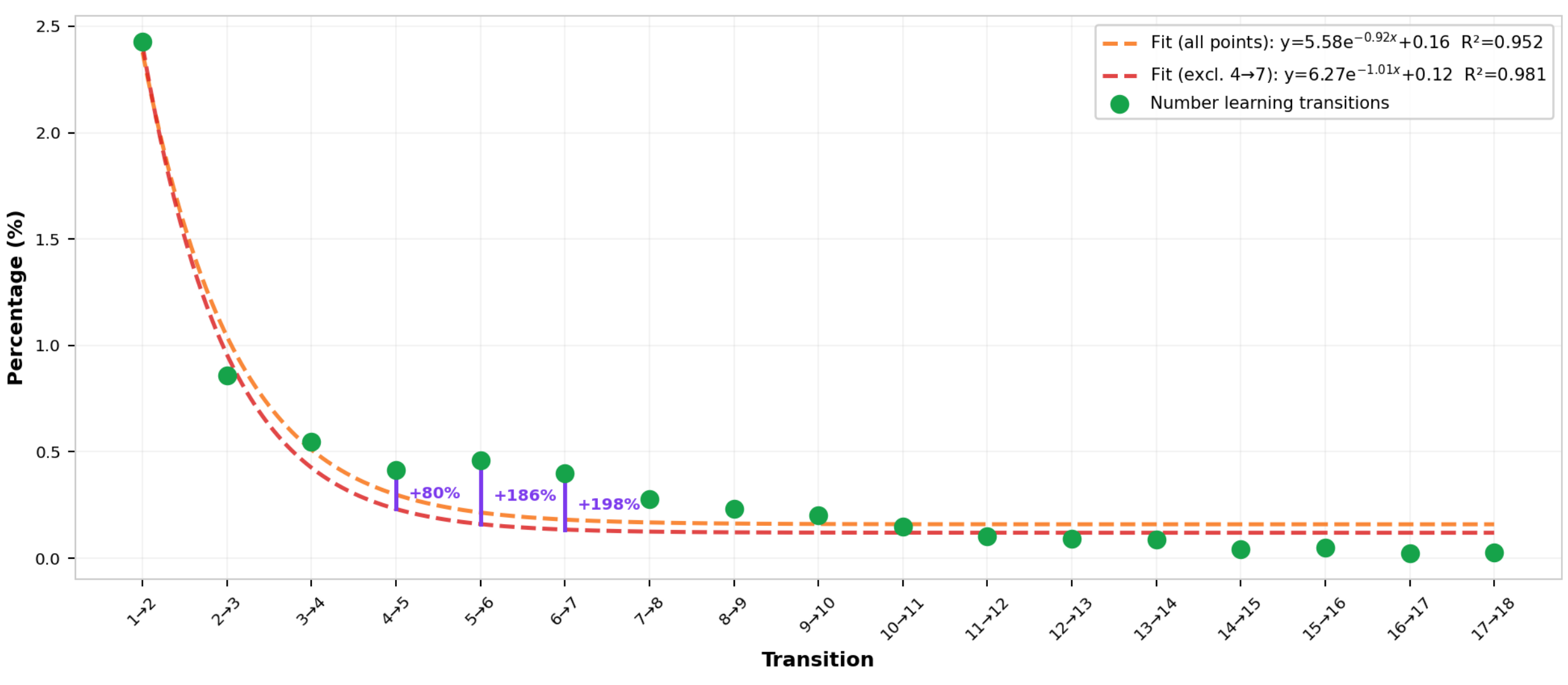

**Fig. 3** Distribution of number learning transitions across parameter space. Percentage of parameter combinations exhibiting learning transitions at each specific stimulation number, from 1→2 through 17→18. Green circles show all number learning transitions. Two exponential decay fits are shown: the orange dashed line fits all points ($y = 5.58e^{(-0.92x)} + 0.16$, $R^2$=0.952) and the red dashed line fits with 4→5, 5→6, and 6→7 excluded ($y = 6.27e^{(-1.01x)} + 0.12$, $R^2$=0.981). Purple residual bars indicate the deviation of these three anomalous transitions above the exclusion fit. The improved goodness of fit when excluding these points ($R^2$=0.981 vs 0.952) confirms that the 4→7 range represents a genuine anomaly rather than an artifact of the fitting procedure. Each data point represents the fraction of all tested parameter combinations (n ≈ 220,000) that exhibited a clean step-change transition at that specific stimulation number.

### *3.3 Number learning transitions follow an exponential decay pattern with anomalous peaks in the 4-7 range*

To determine whether certain stimulation numbers are more conducive to learning than others, we calculated the percentage of parameter combinations exhibiting transitions at each specific stimulation number from 1⟶2 through 17⟶18. We observed that transition frequencies generally decreased exponentially with increasing stimulation number. We fitted two exponential decay curves: one to all data points ($y = 5.58e^{(-0.92x)} + 0.16$, $R^2=0.952$) and one excluding the three anomalous transitions at 4⟶5, 5⟶6, and 6⟶7 ($y = 6.27e^{(-1.01x)} + 0.12$, $R^2=0.981$). The improved goodness of fit when excluding these three points confirms that they represent genuine deviations from the expected monotonic decay rather than an artifact of the fitting procedure (→ Fig. 3). The 1⟶2 transition was most common (2.43% of parameter combinations), while transitions beyond 10⟶11 each occurred in less than 0.15% of cases. Therefore, we conclude that while learning difficulty generally increases exponentially with stimulation number, the system exhibits enhanced learnability specifically in the 4⟶7 stimulation range, suggesting that certain numerical patterns may be intrinsically more accessible to ecological dynamics than the monotonic decay model would predict.

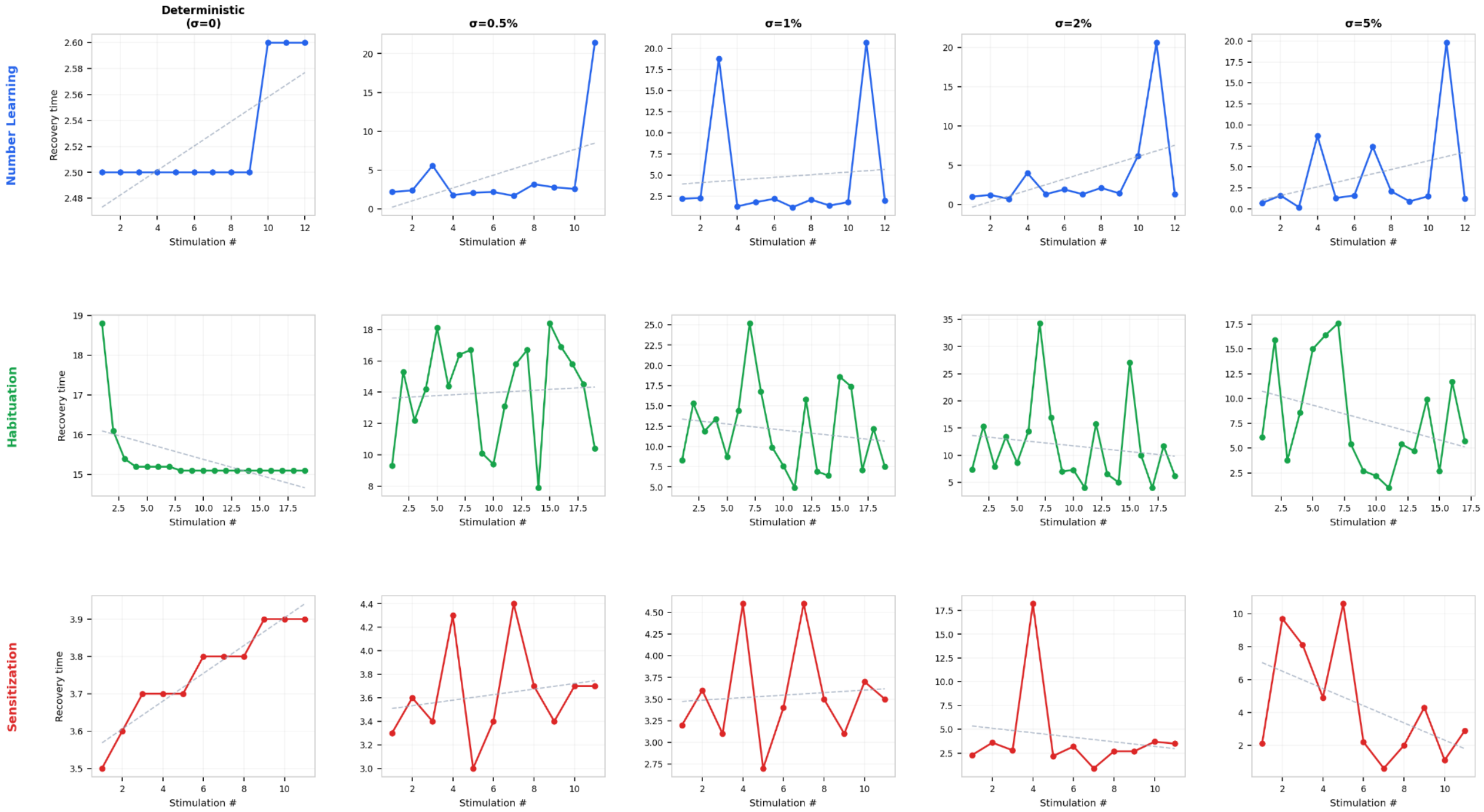

**Fig. 4** Noise robustness of learning phenomena under stochastic perturbations. Recovery time trajectories for three representative parameter sets under increasing levels of Gaussian noise added to prey population dynamics (σ = 0%, 0.5%, 1%, 2%, and 5% of $x_0$ = 25). Rows correspond to number learning (PS=5.0, PF=550, $r_x$=0.70, $r_y$=0.65, $\alpha_{xy}$=0.80, $\alpha_{yx}$=1.30), habituation (PS=23.0, PF=350, $r_x$=0.70, $r_y$=0.60, $\alpha_{xy}$=0.90, $\alpha_{yx}$=0.80), and sensitization (PS=13.0, PF=600, $r_x$=1.40, $r_y$=0.65, $\alpha_{xy}$=0.80, $\alpha_{yx}$=1.20). Dashed grey lines show linear trend fits to recovery time trajectories. Habituation and sensitization preserve their characteristic monotonic trends across all noise levels, while the discrete step-change transition defining number learning is no longer detectable at σ = 0.5%.

## *3.4 Noise robustness of learning phenomena*

To test whether the observed learning phenomena survive stochastic perturbations, we introduced Gaussian noise into the prey population dynamics at increasing levels (σ = 0%, 0.5%, 1%, 2%, and 5% of the initial population size $x_0$ = 25) across three representative parameter sets corresponding to number learning, habituation, and sensitization. In the deterministic case, all three response categories were cleanly detected. Habituation and sensitization were robust to noise across all levels tested, with recovery time trajectories preserving their characteristic monotonic trends even at σ = 5%. Number learning, however, did not survive even the lowest noise level tested (σ = 0.5%), with the discrete step-change transition no longer detectable. This indicates that while the broader trend-based behaviors are robust to stochastic perturbations, the number learning phenomenon relies on precise recovery dynamics that are disrupted by small fluctuations in prey population size (→ Fig. 4).

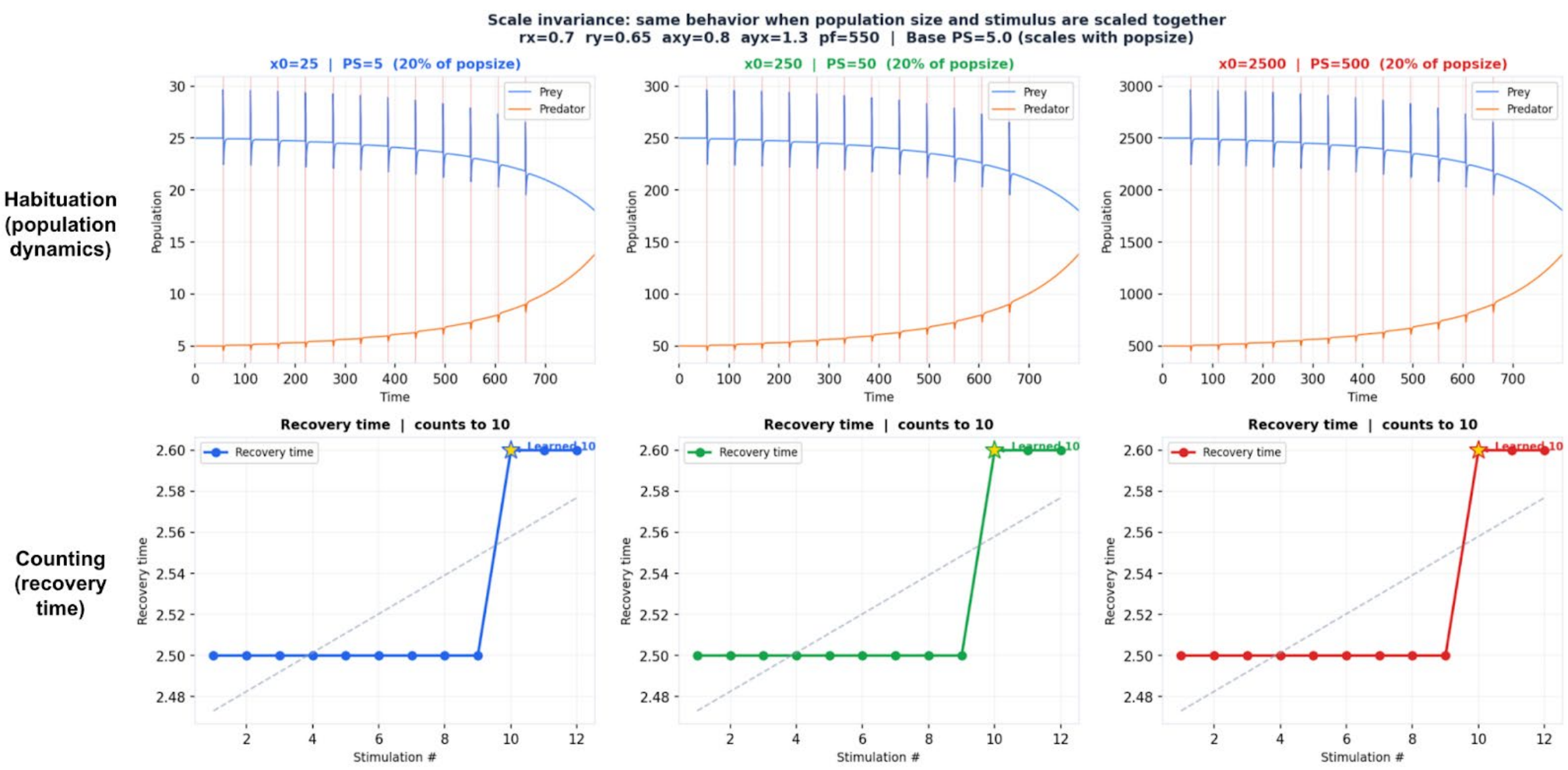

**Fig.** 5 Scale invariance of counting behavior under proportional stimulus scaling. Top row (A–C): population dynamics for x0=25, 250, and 2500 with stimulus scaled proportionally (PS=20% of popsize). All three populations exhibit identical oscillation shapes. Bottom row: recovery time trajectories for each population size, all showing the same 9→10 learning transition at the same stimulation number, confirming scale invariance.

## *3.5 Learning phenomena are scale-invariant*

In order to determine whether the learning behaviors we observed were specific to the population sizes used in our initial parameter sweep, we tested whether habituation and counting transitions would persist when population size and stimulus magnitude were scaled together proportionally. We ran simulations for three population sizes (x0=25, 250, and 2500) with stimulus size set to 20% of the population size in each case, and compared population dynamics and recovery time trajectories across all three conditions. As shown in → Fig. 5, all three population sizes exhibited identical oscillation shapes in their population dynamics, and all three recovery time trajectories showed the same 9→10 counting transition at the same stimulation number. This confirms that the learning phenomena we report are scale-invariant: what matters is the stimulus as a fraction of population size, not its absolute value. The same result held for habituation, with all three scaled populations showing identical habituation slopes when stimulus was expressed as a percentage of population size. These findings indicate that the learning capacity of the system is an intrinsic property of the dynamical interactions rather than an artifact of the specific population sizes chosen for our parameter sweep.

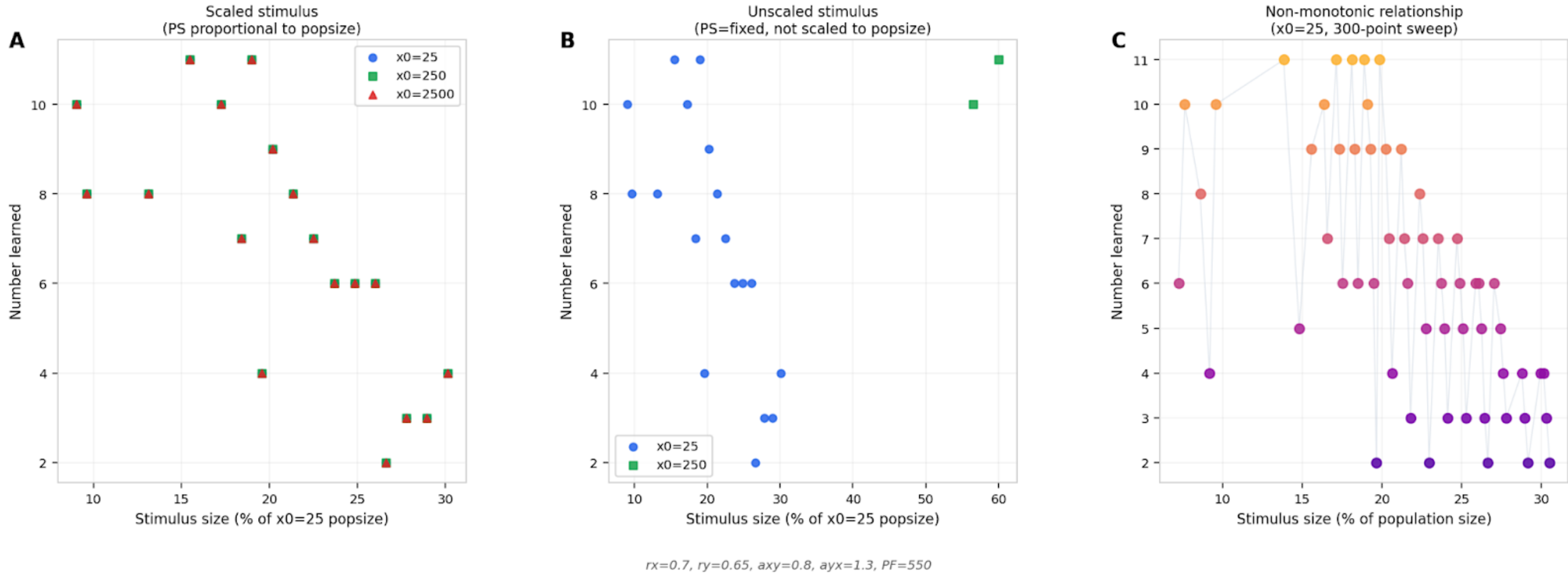

**Fig. 6** Stimulus-size dependence of counting behavior. (A) When stimulus scales with population size, all three population sizes overlap, producing identical number-learned outcomes. (B) When stimulus is held fixed, only the baseline population (x0=25) learns across the full range; larger populations are unresponsive to the same absolute stimulus. (C) High-resolution 300-point sweep for x0=25 reveals a strongly non-monotonic relationship between stimulus strength and learned number, with the system jumping irregularly between counts of 2–11 with no smooth trend.

### *3.6 Counting behavior exhibits scale invariance only when stimulus scales proportionally with population size*

To determine whether scale invariance holds under different stimulation regimes, we tested counting behavior across three population sizes ($x_0$ = 25, 100, and 1000) under two conditions: stimulus scaling proportionally with population size, and stimulus held fixed at the same absolute value across all population sizes. When stimulus size scaled proportionally, all three population sizes produced identical learned number outcomes across the full range of tested stimulus strengths, confirming scale invariance of counting behavior (→ Fig. 6A). However, when stimulus size was held fixed, only the baseline population ($x_0$ = 25) exhibited counting across the full range; larger populations were entirely unresponsive to the same absolute stimulus, demonstrating that scale invariance is contingent on proportional stimulation rather than an intrinsic property of the system independent of stimulus scaling (→ Fig. 6B).

A high-resolution 300-point sweep of stimulus strength for $x_0$ = 25 revealed a strongly non-monotonic relationship between stimulus strength and learned number, with the system jumping irregularly between counts of 2 and 11 with no smooth trend (→ Fig. 6C). This non-monotonic structure indicates that small changes in stimulus strength can produce large and unpredictable changes in which number the system learns, suggesting that the mapping from stimulus strength to learned number reflects complex sensitivity to the precise dynamical regime rather than a simple monotonic tuning relationship.

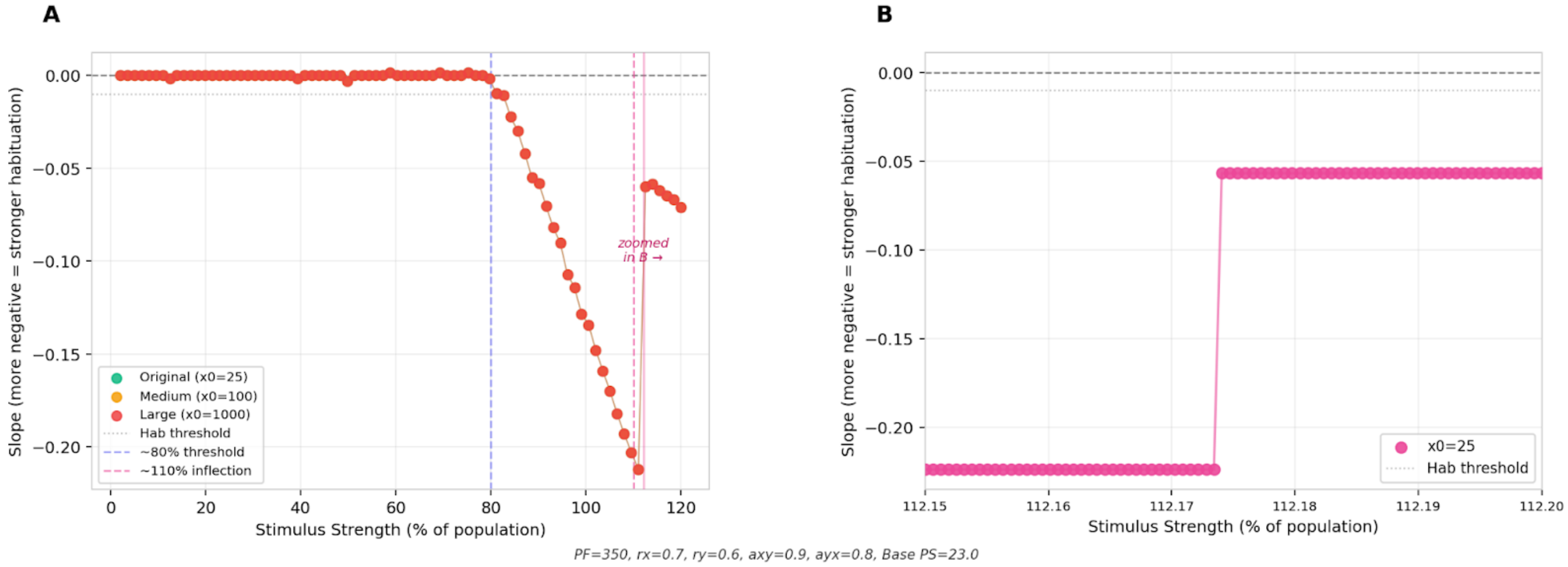

**Fig. 7** Stimulus strength dependence of habituation response. (A) Habituation slope (linear regression of recovery time across stimulations) as a function of stimulus strength (% of population size), swept from 0–120% for three population sizes (x0=25, 100, 1000). All three populations overlap, confirming scale invariance. Below ~80% of population size, no habituation occurs (slope ≈ 0); above this threshold habituation strengthens sharply, with a dip and partial recovery visible near 112%. (B) High-resolution zoom into the 112.15–112.20% region for x0=25, revealing that the sharp transition at this inflection point persists even at fine resolution with no smooth intermediate structure, suggesting a discrete dynamical boundary rather than a gradual change in system behavior. Parameters: PF=350, rx=0.7, ry=0.6, axy=0.9, ayx=0.8, base PS=23.0.

## *3.7 Habituation response is sensitive to stimulus strength*

To characterize how stimulus magnitude influences habituation, we swept stimulus strength from 0 to 120% of population size across three population sizes (x0=25, 100, and 1000) and measured the habituation slope for each condition. As shown in → Fig. 7A, all three population sizes produced identical slope trajectories when stimulus was expressed as a percentage of population size, confirming that habituation is scale-invariant in the same way as counting behavior. Below approximately 80% of population size, no habituation occurred and the slope remained at zero. Above this threshold, habituation strengthened sharply with increasing stimulus, but with a notable non-monotonicity: the slope reached a local minimum near 112% before partially recovering, producing a dip and partial recovery rather than a smooth monotonic descent. To investigate whether this transition was continuous or discrete, we zoomed into the 112.15 to 112.20% region at high resolution (→ Fig. 7B). Even at this fine scale, the transition remained sharp with no smooth intermediate structure visible, suggesting that the inflection point reflects a discrete dynamical boundary rather than a gradual change in system behavior. These findings demonstrate that habituation strength is highly sensitive to stimulus magnitude and that the relationship between stimulus strength and adaptive response is more complex than a simple monotonic trend.

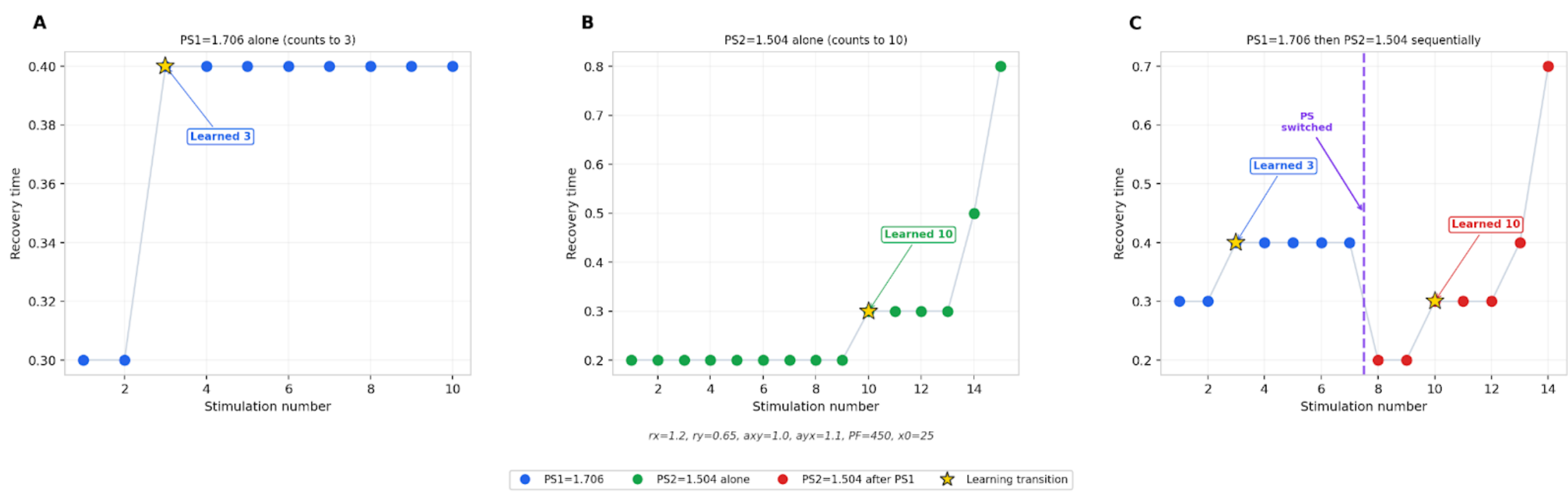

**Fig. 8** Counting behavior does not exhibit meta-memory across sequential pulse size changes. Recovery time trajectories for three conditions using the same system parameters throughout, varying only pulse size (PS). (A) Pulse size PS1=1.706 alone: the system counts to 3, transitioning at stimulation 3. (B) Pulse size PS2=1.504 alone: the system counts to 10, transitioning at stimulation 10. (C) PS1 followed by PS2 sequentially: the system first counts to 3 under PS1, then after switching to PS2 at stimulation 7, the recovery time drops transiently before settling and transitioning at stimulation 10, the same as in panel B. The identical transition point in panels B and C indicates that prior counting experience does not affect subsequent counting behavior, suggesting the system does not retain meta-memory across pulse size changes. Parameters: rx=1.2, ry=0.65, axy=1.0, ayx=1.1, PF=450, x0=25.

### *3.8 Counting behavior does not exhibit meta-memory*

To determine whether prior counting experience affects subsequent counting behavior, we ran a sequential pulse size experiment using a parameter set where two distinct pulse sizes produce clean isolated transitions in isolation. Pulse size PS1=1.706 produced a transition at stimulation 3 when applied alone (→ Fig. 8A), and pulse size PS2=1.504 produced a transition at stimulation 10 when applied alone (→ Fig. 8B). In the sequential experiment (→ Fig. 8C), we applied PS1 until the system settled into its learned state, then switched to PS2 mid-run and continued stimulating on one continuous recovery time trajectory. After the switch, the recovery time dropped transiently before settling, which we attribute to the system still being in the dynamical state shaped by PS1 at the moment of the switch. The system then transitioned at stimulation 10 under PS2, identical to the transition point observed in panel B. This result indicates that prior counting to 3 under PS1 did not affect when or how the system subsequently counted under PS2. Therefore, we conclude that the counting behavior in this system does not retain meta-memory across pulse size changes, and that each counting behavior is determined entirely by current stimulus conditions rather than prior stimulation history. We note that this experiment was conducted on a single example and that clean sequential transitions were rare across parameter combinations tested, consistent with this being an uncommon but observable phenomenon.

### *3.9 Ecological interaction strength dominates parameter space organization of adaptive responses*

To visualize how system parameters determine adaptive response patterns, we performed dimensionality reduction on the six-dimensional parameter space (pulse size, pulse frequency, prey growth rate, predator growth rate, and both interspecific competition coefficients) using Uniform Manifold Approximation and Projection (UMAP). We observed clear spatial separation between habituation and sensitization regions in the two-dimensional embedding, with multiple distinct clusters within each response category revealing subtypes with different parameter configurations (→ Fig. 9). The sensitization region (n=65,944 points) formed numerous well-separated clusters distributed throughout the parameter space, while habituation (n=7,021 points) occupied a more compact, localized region. Analysis of parameter contributions using Pearson correlation with UMAP dimensions revealed that interspecific competition coefficients ($\alpha_{xy}$: r = -0.945 with UMAP dimension 1; $\alpha_{yx}$: r = 0.596 with UMAP dimension 1, r = 0.761 with UMAP dimension 2) and pulse size (r = -0.816 with UMAP dimension 2) were the dominant factors organizing the parameter space, collectively explaining the spatial structure. In contrast, pulse frequency and intrinsic growth rates contributed minimally ($|r| < 0.11$ for all), indicating that temporal parameters have negligible influence on response category boundaries. Therefore, we conclude that adaptive response patterns are primarily determined by ecological interaction strengths (how strongly species compete) and perturbation magnitude rather than by temporal parameters (when perturbations occur) or intrinsic growth rates (how fast populations grow in isolation).

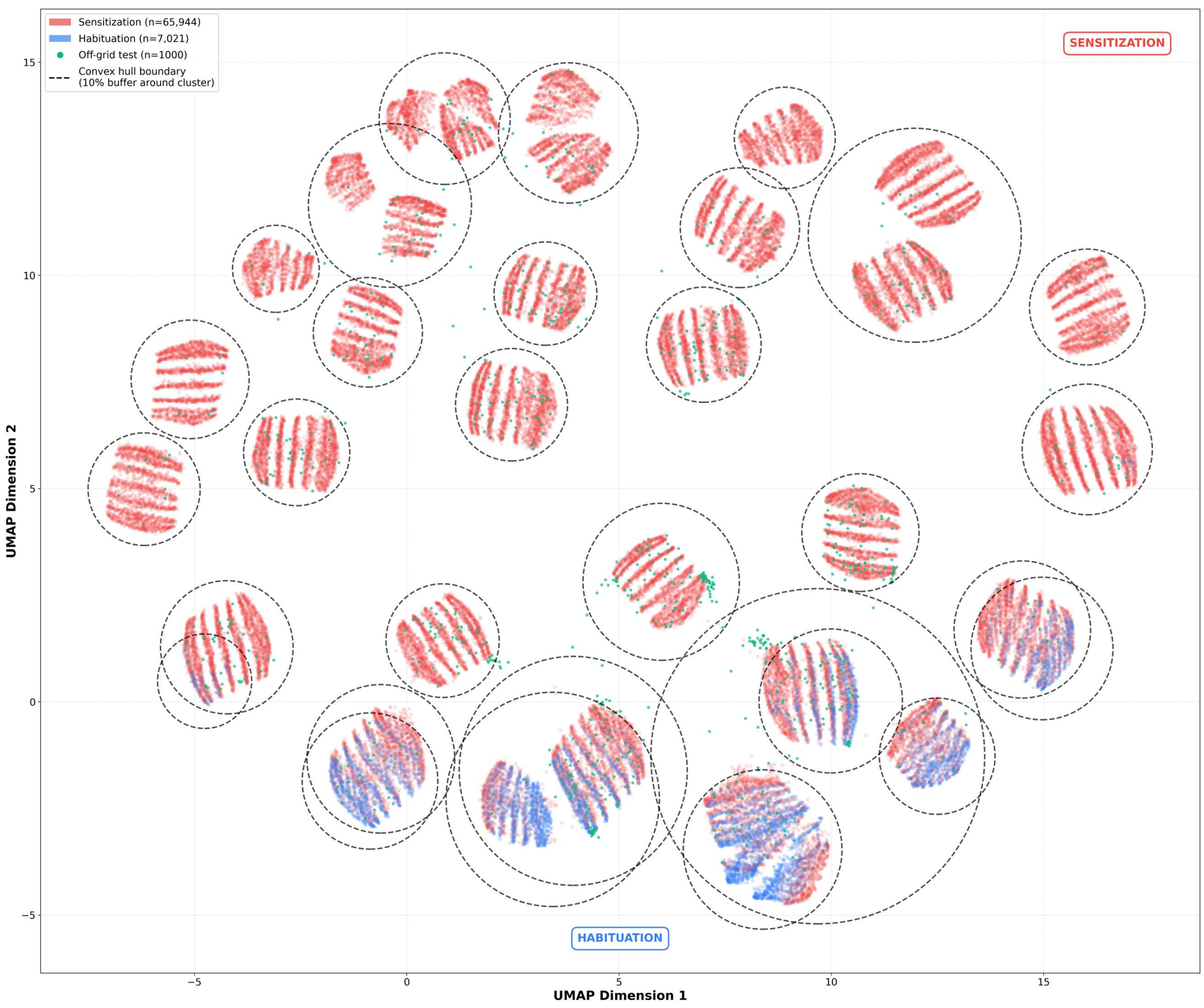

**Fig. 9** UMAP visualization reveals distinct parameter space organization for habituation and sensitization responses. Uniform Manifold Approximation and Projection (UMAP) embedding of the six-dimensional parameter space (pulse size, pulse frequency, prey and predator growth rates, and interspecific competition coefficients) for all simulations classified as habituation (n=7,021, blue) or sensitization (n=65,944, red). Dashed circles represent convex hulls (10% buffer) that delineate spatial clustering within each response category in the two-dimensional embedding, revealing multiple distinct "subtypes" of habituation and sensitization in different parameter regions. Green points represent test parameter combinations (n=1,000) sampled from regions of parameter space not included in the training data. Testing on these 'held-out' parameters validates that the model can predict recovery time patterns for entirely new parameter combinations, demonstrating its generalization ability beyond the training set. Of these test points, 98.1% fell within the boundaries of their predicted response category (0 in habituation regions, 981 in sensitization regions), with only 19 points (1.9%) falling outside all category boundaries, demonstrating robust parameter space partitioning and predictive power of the classification approach.

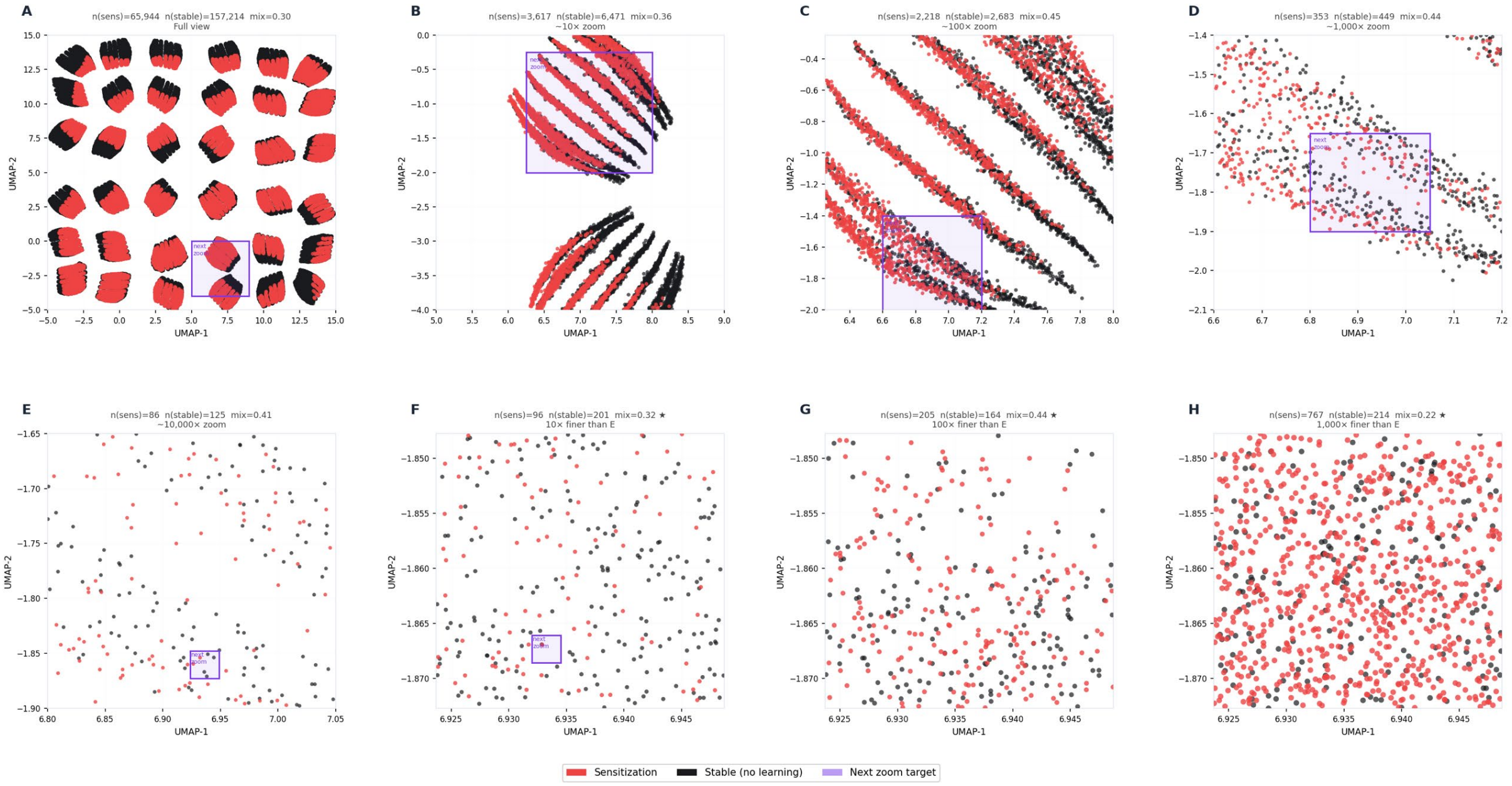

**Fig. 10** Fractal-like boundary structure persists across successive zoom levels in UMAP parameter space. Progressive zoom sequence (A-H) into the UMAP embedding shown in → Fig. 9. Panels A-E show original dataset points at successively finer zoom levels, targeting a region containing both sensitization (red) and stable non-learning (black) parameter combinations. Panels F-H show 46,656 new simulations each, run at progressively finer parameter resolution (10×, 100×, and 1,000× finer than panel E respectively) and projected into the same global UMAP window. Purple rectangles indicate the region targeted in the subsequent panel. Mix scores (fraction of minority class) remain above 0.22 across all panels, confirming that both response categories remain intermixed at every level of magnification rather than resolving into uniform regions. This persistence of mixed structure across panels A through H, spanning approximately five orders of magnitude in parameter space resolution, suggests that the boundary between learning and non-learning parameter regimes has fractal-like fine structure that does not simplify upon closer inspection.

## *3.10 Parameter space boundary exhibits fractal-like structure*

To determine if the parameter space is rugged, we sought to investigate the space between successful parameters at higher resolution. We found that even at successively finer zoom levels into the UMAP embedding, both sensitization and stable non-learning parameter combinations remained intermixed rather than resolving into distinct uniform regions (→ Fig. 10). Starting from the full embedding in panel A and zooming progressively through panels B to F, both response categories persisted at every level of magnification with no zoom level producing a homogeneous region of a single category. This suggests that the boundary between learning and non-learning parameter regimes has fractal-like fine structure that does not simplify upon closer inspection. The persistence of this mixed structure across panels A through H spans approximately five orders of magnitude in parameter resolution: panels A through E cover the original data from the full parameter landscape down to the ~10,000x zoom level, while panels F, G, and H introduce 46,656 new simulations each at 10x, 100x, and 1,000x finer parameter resolution than panel E respectively, all maintaining mix scores above 0.22. This implies that the parameter space is genuinely rugged, with learning-capable and non-learning parameter combinations interspersed at arbitrarily fine scales rather than occupying cleanly separable contiguous regions.

### *3.11 Parameter space clustering enables robust prediction of adaptive responses for novel parameter combinations*

To validate that the observed parameter space organization could predict responses for untested parameter combinations, we sampled 1,000 test points from regions not included in the original parameter sweep, including intermediate values between grid points and boundary regions near the edges of the explored parameter space. We projected these test points into the UMAP embedding (→ Fig. 9) using the trained manifold transformation and assessed whether they fell within the convex hull boundaries (10% buffer) of their predicted response categories determined by proximity to training points. We observed that 981 of 1,000 test points (98.1%) fell within sensitization region boundaries, 0 points fell within habituation boundaries (consistent with habituation's relative rarity and compact spatial extent), and only 19 points (1.9%) fell in intermediate regions outside all category boundaries, likely representing genuinely ambiguous parameter combinations near category transitions. Among the correctly classified test points, the spatial distribution matched the training data, with test points populating the same cluster regions as their training counterparts. Therefore, we conclude that the parameter space exhibits robust partitioning between adaptive response regimes, with the classification approach demonstrating strong predictive power (98.1% accuracy) for entirely novel parameter combinations beyond the training set, validating the biological relevance of the identified parameter space structure.

**Table 1** Parameters from published ecological models tested against our perturbation protocol.

| # | System | Description | $r_x$ | $r_y$ | $\alpha_{xy}$ | $\alpha_{yx}$ | References | Learning Type |
|---|---|---|---|---|---|---|---|---|
| 1 | Lynx-hare (Hudson Bay) | Canonical mammalian predator-prey cycle; Hudson Bay pelt records | 1.00 | 1.00 | 0.035 | 0.010 | [44, 45, 54] | Stable |
| 2 | Lynx-hare (Bayesian fit) | Modern Bayesian inference of L-V parameters from the same time series | 1.00 | 1.00 | 0.040 | 0.020 | [55] | Stable |
| 3 | Gut microbiome (gLV) | Generalized L-V model fitted to mouse intestinal microbiota under antibiotic perturbation | 0.90 | 0.90 | 0.050 | 0.20 | [56] | Stable |

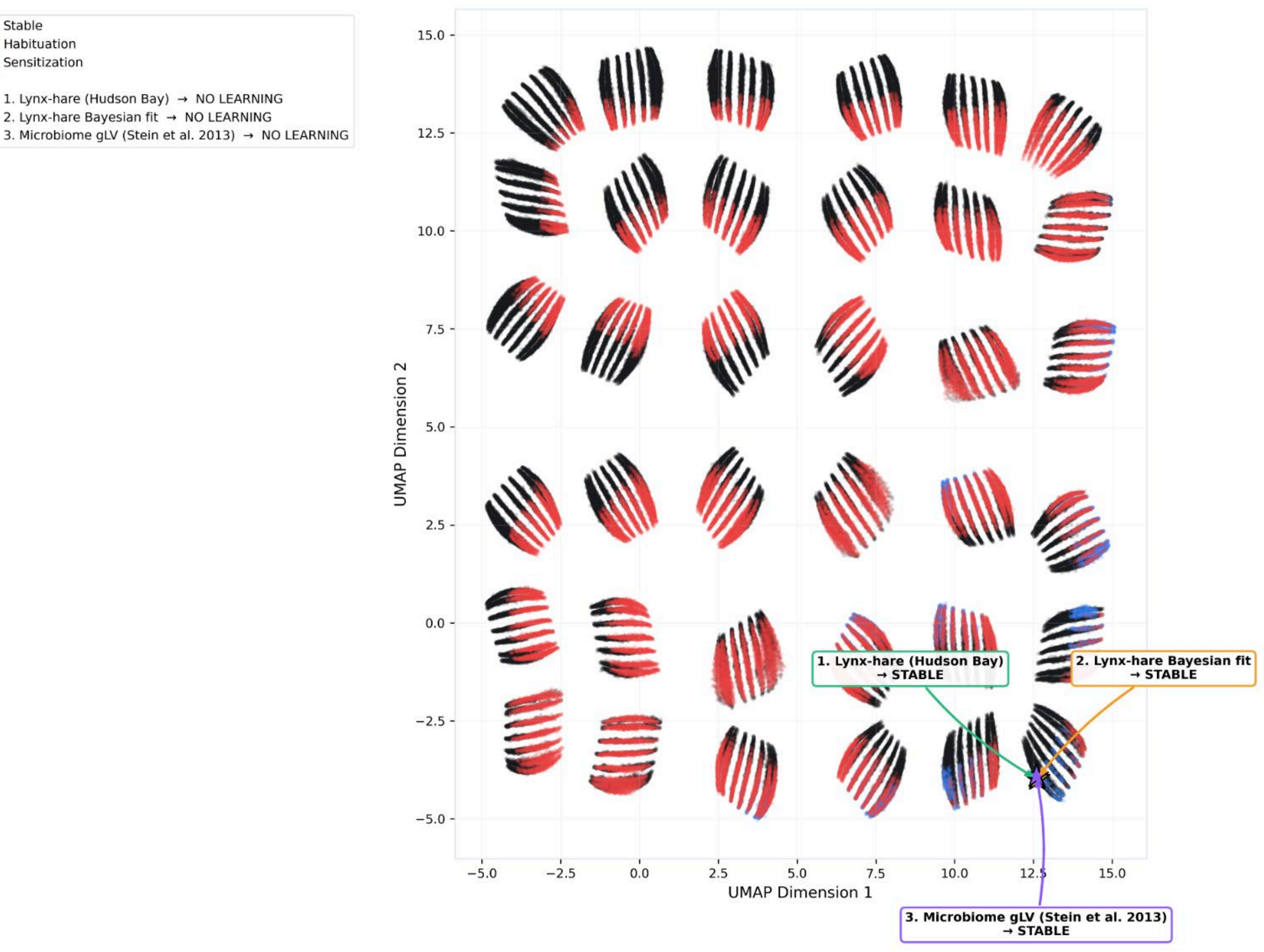

**Fig. 11** Parameters from published ecological models project into stable regions of the UMAP parameter space. UMAP embedding of the full parameter space (same as → Fig. 9), showing all sensitization (red) and stable non-learning (black) parameter combinations from the original sweep. Colored stars indicate representative parameter sets drawn from three published predator-prey and ecological models (see Table 1). All three project into stable, non-learning regions of the embedding. This result indicates that while these real biological systems operate under parameter regimes that produce stable predator-prey dynamics, they do not fall within the learning-competent zones identified in our parameter sweep under the perturbation protocol used here.

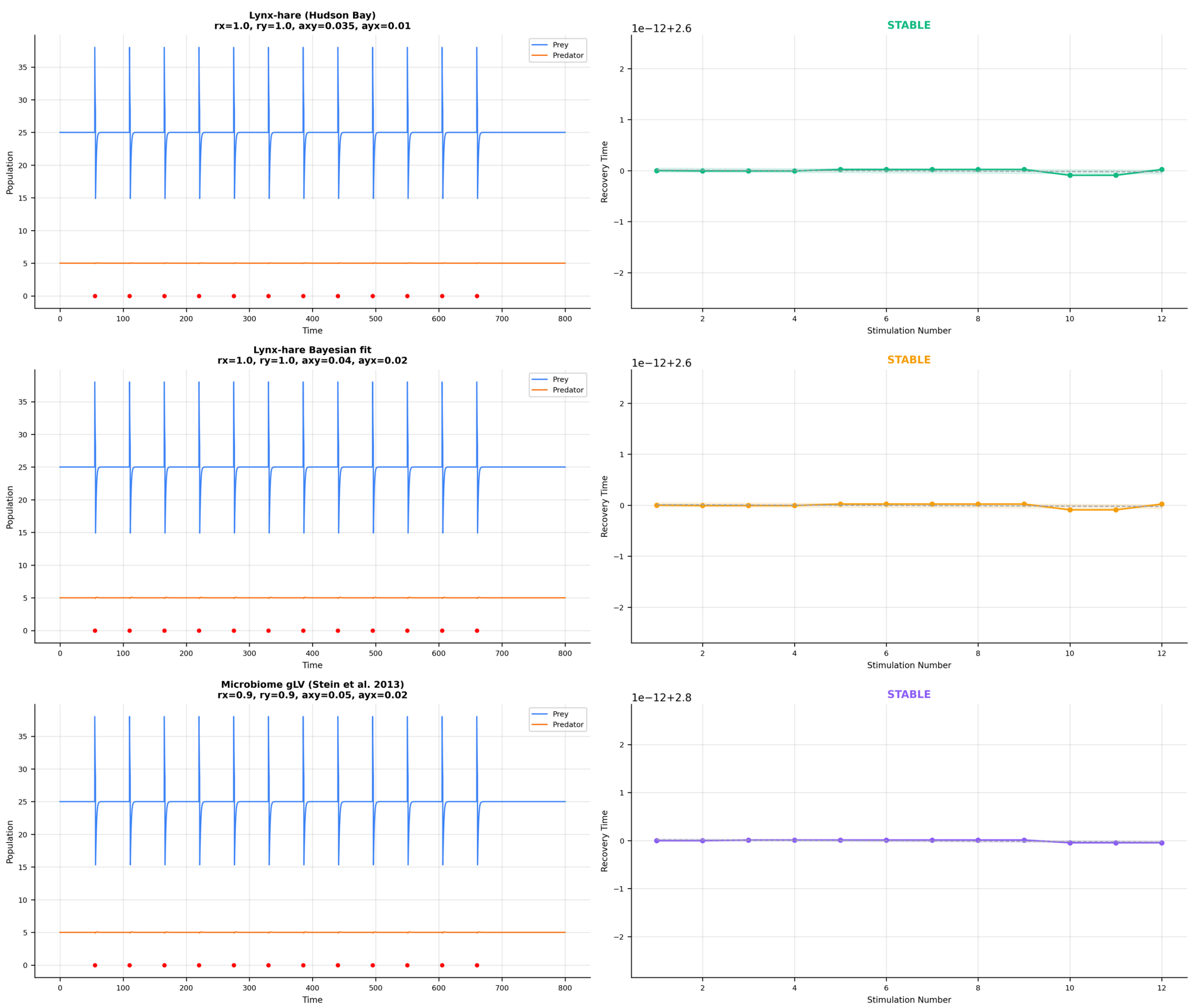

**Fig. 12**. Population dynamics and recovery time trajectories for three published ecological models under the perturbation protocol. Each row corresponds to one system from → Table 1. Left column: prey (blue) and predator (orange) population trajectories over the full simulation, with red dots marking perturbation events. Right column: recovery time at each stimulation number. All three systems exhibit flat recovery time trajectories with no significant monotonic trend, consistent with stable (non-learning) classification. Dashed lines indicate the linear trend ± 1 standard deviation. The stability of recovery times across all systems suggests that parameters characteristic of real predator-prey dynamics as modeled in the literature fall outside the learning-competent parameter regime identified in our sweep, though these systems remain close to that regime in UMAP space.

### *3.12 Parameters from published ecological models fall near but outside learning-competent regions of parameter space*

To assess whether the learning phenomena we identified are biologically relevant, we tested whether parameter sets drawn from published predator-prey studies fall within learning-competent regions of our parameter space. We identified three published models spanning canonical ecological and microbial systems: the lynx-hare system fitted to Hudson Bay data [44, 45, 54], a modern Bayesian inference of the same system [55], and a generalized Lotka-Volterra model of gut microbiome dynamics [56]. For each, we selected representative parameter values converted into our carrying-capacity formulation, simulated the full perturbation protocol, and classified the resulting behavior ( → Table 1). All three parameter sets were classified as stable under our perturbation protocol (→ Fig. 12), and all three projected into stable, non-learning regions of the UMAP embedding (→ Fig. 11). Notably, these points cluster near but outside the learning-competent zones visible in the embedding, suggesting that real biological systems operating under these parameter regimes sit close to the boundary of dynamical complexity required for learning. This raises the possibility that modest changes in ecological interaction strengths, achievable through evolutionary or environmental shifts, could push such systems into learning-competent regimes. The result also illustrates the predictive utility of the parameter space map: given the interaction coefficients of a real system, one can immediately assess its proximity to learning capacity without running a full simulation sweep.

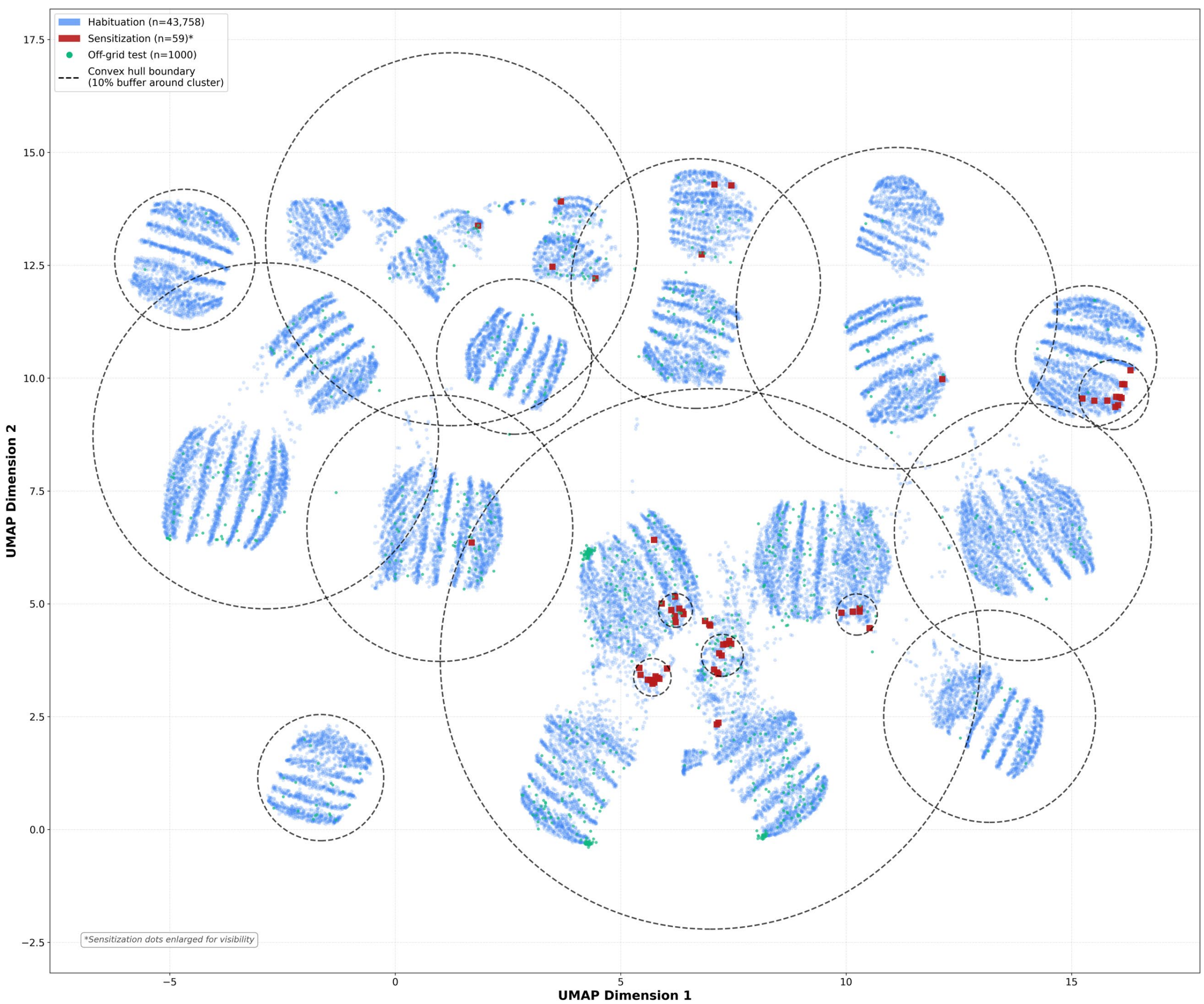

**Fig. 13** Magnitude response patterns occupy distinct parameter space regions compared to recovery time dynamics. UMAP embedding of magnitude response patterns for all simulations classified as habituation (n=43,758, blue circles) or sensitization (n=59, red squares, enlarged for visibility). Dashed circles represent convex hulls delineating spatial clustering within each response category. Unlike recovery time patterns (→ Fig. 8), magnitude responses show habituation dominating across 17 distinct clusters spanning the entire parameter space, with sensitization cases forming a compact central region. Green points represent off-grid test combinations (n=1,000) validating the embedding structure. The magnitude response threshold was set to ±0.005 (compared to ±0.01 for recovery time) to capture weak sensitization signals. This lower threshold reveals that magnitude sensitization is exceptionally rare (0.03% of cases) despite being detectable, suggesting fundamental dynamical constraints on systems exhibiting increasing response amplitudes under repeated perturbation.

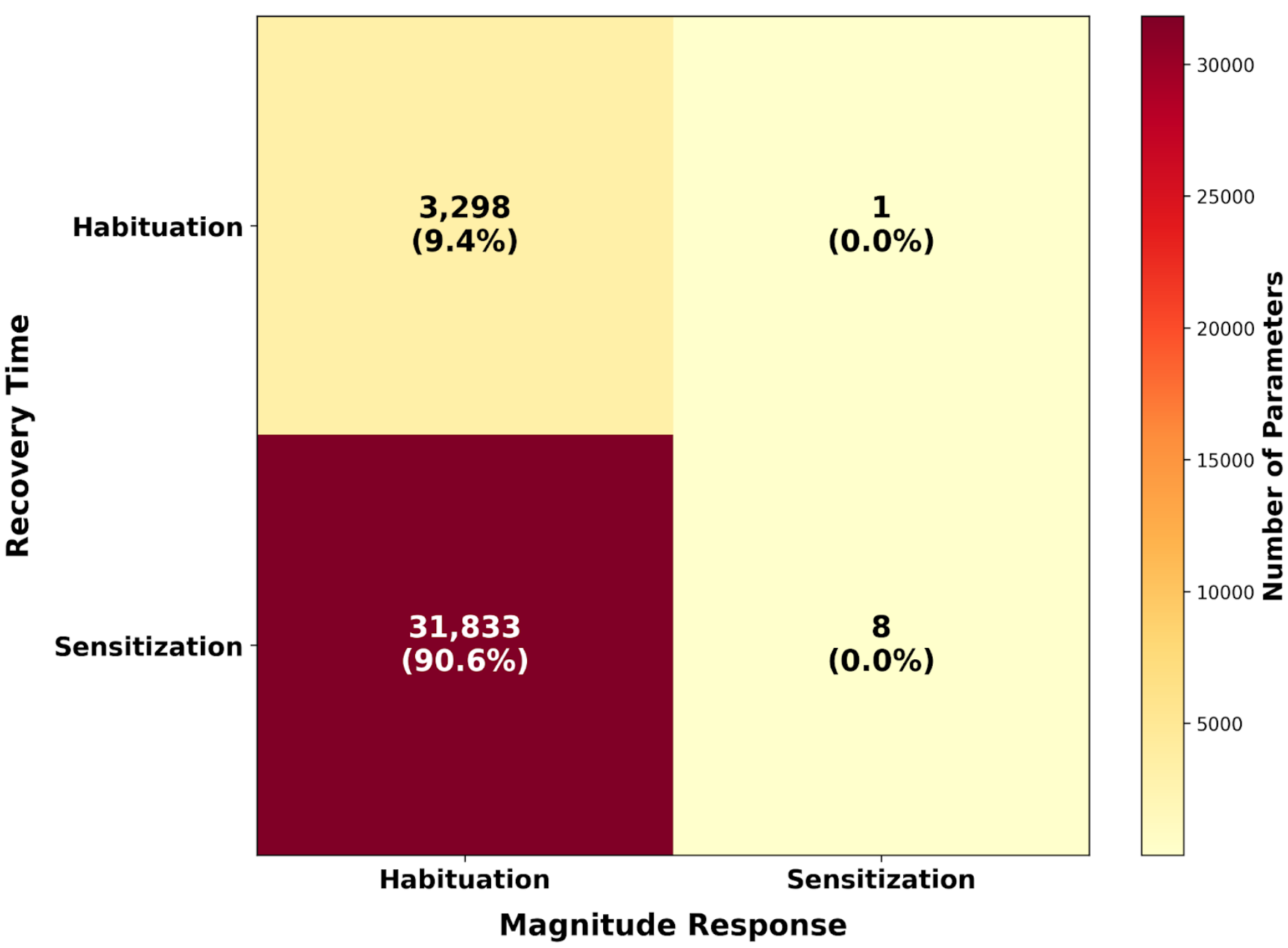

**Fig. 14** Parameter overlap matrix reveals asymmetric coupling between recovery time and magnitude response. Parameter overlap matrix quantifying co-occurrence patterns for the 35,140 parameter combinations exhibiting habituation or sensitization in both metrics. Each cell shows the number and percentage of parameters with a specific combination of recovery time response (rows) and magnitude response (columns). The dominant pattern (90.6%, dark red) shows recovery time sensitization paired with magnitude habituation, indicating that parameters producing slower recovery also constrain equilibrium displacement. The opposite coupling (recovery habituation + magnitude sensitization) occurs in only 0.003% of cases (1 parameter), while both metrics showing the same direction (hab/hab or sens/sens) account for only 9.4% combined. This asymmetry reveals that stable parameter regimes under repeated perturbation preferentially produce dynamics where temporal recovery slows while spatial displacement decreases, likely reflecting fundamental constraints on dynamically stable repeated perturbation responses.

### *3.13 Magnitude response patterns show inverse parameter space organization and asymmetric coupling with recovery time dynamics*

To investigate whether recovery time fully characterizes system adaptation or whether additional metrics reveal complementary information, we analyzed response magnitude (the absolute change in equilibrium prey abundance following each perturbation). Classification of magnitude trajectories across the same 22,176 parameter combinations revealed 43,758 cases (19.73%) exhibiting magnitude habituation, 59 cases (0.03%) exhibiting magnitude sensitization, and the remainder showing stable, insufficient, or no-variation responses (→ Fig. 13). The extreme rarity of magnitude sensitization (200-fold less common than magnitude habituation) contrasts sharply with recovery time patterns, where sensitization was 9-fold more common than habituation, indicating that these metrics capture fundamentally different aspects of system dynamics.

UMAP embedding of magnitude response patterns (→ Fig. 13) revealed spatial organization markedly distinct from recovery time (→ Fig. 9). While recovery time sensitization formed numerous distributed clusters with compact habituation regions (→ Fig. 9), magnitude responses showed the opposite: widespread habituation clusters spanning the entire parameter space with sensitization cases concentrated in a single compact central region. This inverted spatial structure indicates that the parameters governing temporal recovery dynamics differ substantially from those controlling equilibrium displacement magnitude.

To quantify the relationship between these metrics, we identified 35,140 parameter combinations classified as habituation or sensitization in both analyses. Analysis of co-occurrence patterns revealed striking asymmetry: 31,833 cases (90.6%) exhibited recovery time sensitization paired with magnitude habituation, while only 8 cases (0.023%) showed sensitization in both metrics (→ Fig. 14). The remaining cases split between both metrics showing habituation (3,298 cases, 9.4%) and recovery habituation with magnitude sensitization (1 case, 0.003%). This 90.6% predominance of the sensitization-habituation coupling indicates that parameter combinations supporting stable repeated perturbation responses are characterized by a specific dynamical signature: progressively slower recovery paired with progressively smaller equilibrium displacements. The near-absence of dual sensitization (both metrics increasing) likely reflects dynamical instability, as systems exhibiting both increasing recovery times and increasing displacement magnitudes would rapidly fail under continued perturbation.

## 4 Discussion

We found that a simple predator-prey model can exhibit habituation, sensitization, and number learning. The system switches to new recovery states only after experiencing specific numbers of perturbations, ranging from 2 to 18. This happened across thousands of parameter combinations without needing specialized biology, evolution, or explicit memory storage. The math itself creates the ability to learn. Our parameter exploration showed that ecological interaction strengths determine which adaptive responses appear, while timing factors like perturbation frequency barely matter. The clear organization in parameter space let us predict learning in new parameter combinations with high accuracy, confirming these patterns are real features of the system. We also found that two different response metrics capture different aspects of adaptation. Most stable parameter combinations show a specific pattern where temporal recovery slows down while spatial displacement decreases.

### *4.1 Number learning as emergent computation*

We demonstrated that a simple two-species predator-prey model can exhibit three distinct response patterns when subjected to repeated perturbations: habituation (decreasing recovery times), sensitization (increasing recovery times), and discrete number learning (abrupt transitions to new recovery baselines at specific stimulation counts). Most strikingly, we identified a novel form of discrete number learning in which systems transition to new recovery baselines only after experiencing specific counts of perturbations (ranging from 2 to 18). These findings suggest that ecological dynamics could exhibit unexpected computational properties, including numerical pattern recognition, without requiring neural substrates or explicit memory mechanisms.

### *4.2 Memory does not require parameter fine-tuning*

The observation that 3.1% of tested parameter combinations exhibited number learning (6,821 out of 220,000) demonstrates that this computational capacity does not require precise parameter fine-tuning. Similarly, sensitization occurred in 30.0% of tested combinations (65,944 out of 220,000) and habituation in 3.2% (7,021 out of 220,000), further demonstrating that these adaptive capacities arise broadly across parameter space rather than requiring specialized conditions. In other words, the ability to react after a specific number of stimuli (a primitive kind of counting) is not a needle-in-a-haystack rare property that needs to be carefully engineered into a system, or one that appears only after many generations of evolutionary selection. It is an intrinsic property of this class of systems and can be expected to be found in even a relatively small set of populations with randomly-chosen parameters, akin to a kind of fine-tuning that has been suggested for physical and cosmological parameters that favor appearance of intelligence (life) [57-60].

### 4.3 Effective dimensionality and learning map of the parameter space

We included PS and PF in our study of the parameter space since we did not know in advance that they wouldn't matter to the outcomes. The effective dimensionality of the system is 5, because Kx and Ky are fully determined (there are no free parameters in K once one sets the competition coefficients and initial conditions). Having scanned the space, we found a very unique and interesting shape (Figures 9,10,11,13) along which specific learning parameters lie. This shape appears to be an intrinsic property of the behavior of these mathematical objects under specific bio-inspired styles of perturbation.

### 4.4 Recovery time and magnitude response reveal asymmetric dynamical coupling

The striking predominance of recovery time sensitization paired with magnitude habituation (90.6% of overlapping parameters) reveals a fundamental asymmetry in how ecological systems respond to repeated perturbations. This coupling pattern emerges not from optimization or selection, but as a natural consequence of the underlying dynamical equations. In parameter combinations that support stable dynamics under repeated forcing, the mathematical structure governing recovery dynamics simultaneously influences equilibrium displacement: parameters that slow temporal recovery also constrain spatial deviation from baseline.

The extreme rarity of dual sensitization (0.023%, 8 cases) supports this interpretation. Parameter combinations producing both increasing recovery times and increasing displacement magnitudes would generate progressively larger deviations requiring progressively longer resolution periods (a trajectory toward dynamical failure rather than stable adaptation). The asymmetric coupling observed here may therefore reflect a fundamental constraint: systems maintaining stability under repeated environmental perturbations can exhibit temporal sluggishness (slower recovery) or spatial deviation (larger displacements), but rarely both simultaneously.

The two metrics capture distinct but complementary dynamical properties. Recovery time quantifies temporal responsiveness (the rate at which restoring forces return the system to equilibrium following displacement). Magnitude quantifies spatial displacement (the extent to which perturbations shift equilibrium positions). Their coupling reveals that in most stable parameter combinations, repeated perturbations weaken temporal restoring forces while strengthening spatial stability constraints. This suggests that the dominant mode of adaptation in this system involves accepting slower recovery dynamics in exchange for maintaining equilibrium positions close to baseline.

The distinct spatial organizations observed in UMAP embeddings for the two metrics further demonstrate their complementary nature. While both partition parameter space into habituation and sensitization patterns, the inverted cluster structures (distributed sensitization for recovery time versus concentrated sensitization for magnitude) indicate that different subsets of the six system parameters govern these two aspects of adaptation. Future mathematical analysis investigating how specific parameter combinations produce this coupling may reveal deeper principles governing stable responses to repeated forcing in dynamical systems.

### 4.5 Broader issues of unconventional memory effects

A few comments are warranted about the interpretation of memory effects in ecosystems and beyond

First, what counts as a "learning event"? We used standard behavior science definitions of habituation, sensitization, etc. which refer to predictable, monotonic changes in response strength or recovery duration following repeated stimuli. However, we chose the strength of the stimulus and the sensitivity with which we detected increase/decrease of response, by visual inspection of a number of preliminary experiments. This is no different than in behavioral studies of biological memory. For example, Pavlovian experiments do not set “salivation threshold” at 1 Liter or 1 nanoLiter, since both would fail to reveal the interesting phenomenon of associative conditioning. In all such cases, one must define the criteria that are appropriate to the behavior of the system. Nevertheless, we do not know yet what sensitivity of inputs and outputs in such systems might be of ecological significance, on Earth and perhaps elsewhere.

Second, was there a Null Hypothesis? There was – that such systems would show no phenomena matching the behavioral definition of habituation, sensitization, or counting. A priori, before beginning our experiments, there were 3 possibilities: 1) no such systems show habituation, sensitization, etc. behavior, 2) all such systems show these effects, or, 3) only some (for certain parameters) show them. Our hypothesis was that some would, but we didn't know whether any would, or how rare it would be in the space of possible parameter values. It would be falsified, just like it could have been for memory in gene-regulatory models [46, 47], if none of them showed it. These kinds of equation systems could have proven to be too simple to show that kind of memory, or too chaotic; habituation/sensitization behavior is a very specific kind of response to repeated stimuli and many systems don't show it, being either too inflexible to do anything different after repeated same stimulus, or too responsive, shooting off into states that make it impossible to see slow up/down trends from repeated, pulsed stimuli. Note that this is not the same thing as stability, or mere responsiveness. If the effects were chaotic (or just non-monotonic), or anything other than the traditional behavioral curve for habituation/sensitization, we would not have counted them as examples of learning. However, it should be pointed out that there is no guarantee that current state of behavioral science has captured all interesting and relevant learning schemas, and it's possible that the standard taxonomy of learning [2, 61] will have to be expanded as research in active matter [22, 56] and diverse intelligence proceeds.

Third, these phenomena can also be described in the language of bifurcation theory. Regarding how the phenomena map onto known bifurcation types, the "number learning" step-transitions closely resemble what happens when a driven nonlinear system accumulates perturbations and crosses a basin boundary or hits a saddle-node bifurcation. The system sits in one attractor basin. Repeated perturbations shift the state. At some critical count it crosses into a new basin with different recovery dynamics. That is similar to the step-change and the specificity of the count that we observed. Habituation and sensitization are harder to map exactly, but monotonic changes in recovery time under repeated forcing are not uncommon in driven oscillators. However, while dynamical systems researchers are well familiar with basin deformation, saddle-node bifurcations under slow drift, and other such phenomena, repeated stimuli to achieve stable changes of response have not to our knowledge been sufficiently exploited. We argue the behavioral science framing is complementary and reveals structure that the dynamical systems framing alone does not highlight. One valuable thing about conceptual maps from different fields is that they can suggest different approaches and experiments to be done – they lead to new roadmaps; this is different from simply noting, after interesting effects are observed, that they are not inconsistent with standard paradigms. The applicability of learning paradigms in systems that are also describable by dynamical systems language is no more (or less) surprising than the existence of conventional learning in brain tissue also in principle describable by mechanical subatomic chemistry. The advantages of using this approach in ecology and beyond are exactly the same reasons for which we use behavioral science to relate to animals and other humans, not the frameworks of particle physics which surely applies to their materials [62, 63].

## 5 Future Work

We focused on number learning, habituation, and sensitization, but ecological systems might be capable of other types of learning. Associative learning could emerge if two different types of perturbations were used, and tested to see whether the system learns to predict one from the other. Anticipatory behavior might occur if predictable perturbation patterns lead to compensatory or responsive dynamics occur when a stimulus is skipped. Likewise extinction learning, where a learned response disappears with prolonged exposure, could be tested to determine whether systems can relearn after extinction. Each type would tell us something different about what computation and proto-cognitive phenomena are possible in these dynamics.

It is not known whether evolution exploits these effects in real ecosystems (or perhaps, avoids them); while such experiments might be impractical in the large scale, these effects can and should be sought in a wide range of tractable dynamical systems. Microbial predator-prey systems like bacteria-phage are good candidates for biological validation because they have short generation times and can be easily stimulated in the lab. However, our findings are not really about ecological populations *per se*. While our analysis was motivated by an interest in learning phenomena in very large-scale systems, it should be noted that there is nothing here that is specifically tied to ecology (or even biology). Lotka-Volterra equations have the same form as many other natural systems. Any system that is well-described by the equations we simulated would be expected to show the memory effects; for example, chemical reaction networks like the Belousov-Zhabotinsky oscillator and epidemic dynamics (SIS and SIR models) all use similar coupled differential equations and could be tested in silico and in vitro.

Neural networks might show similar patterns since they also involve coupled populations with feedback. Power grids and traffic flow use comparable equations too. The potential prevalence of learning in diverse media, regardless of whether it is biological or not, is a feature, not a bug. In terms of practical uses, finding such phenomena in in areas of physiology, ecology, meteorology, and many other domains could help explain failure modes and lead to new applications.

We argue that these and other protocols in behavioral science should be checked in numerous unconventional domains. If number learning appears across multiple domains, it would point to universal mathematical principles for how dynamical systems process temporal information. Finding the minimal mathematical requirements for number learning would connect our work to fundamental dynamical systems theory, as well as provide a necessary grounding for unification of concepts in diverse intelligence across the complexity spectrum.

## 6 Conclusion

These results show that learning and memory can come from dynamical interactions alone. Finding numerical pattern recognition in ecological equations suggests computation might be way more common in nature than we thought. Since the ability comes from mathematical structure and not biological specifics, similar computational properties could show up in other coupled dynamical systems, from molecular networks to social groups. This means evolution and engineering work with systems that already have computational abilities built in. This opens up questions about what other types of computation naturally emerge from dynamics and how mathematical structure enables information processing across different scales.

## Acknowledgments

We thank Hananel Hazan for his very helpful comments on an early version of the manuscript, and Josh Bongard and Richard Watson for many helpful discussions of relevant issues. We thank Eugene Jhong and The Fries Family Foundation for their support.

## References

[1] Thompson, R. F. & Spencer, W. A. Habituation: A model phenomenon for the study of neuronal substrates of behavior. *Psychol. Rev.* **73**, 16–43 (1966).

[2] Rankin, C. H. *et al.* Habituation revisited: an updated and revised description of the behavioral characteristics of habituation. *Neurobiol. Learn. Mem.* **92**, 135–138 (2009).

[3] Kaygisiz, K. & Ulijn, R. V. Can Molecular Systems Learn? https://doi.org/10.1002/syst.202400075 doi:10.1002/syst.202400075.

[4] Reber, A. S. & Baluška, F. Cognition in some surprising places. *Biochem. Biophys. Res. Commun.* **564**, 150–157 (2021).

[5] Baluška, F. & Levin, M. On Having No Head: Cognition throughout Biological Systems. *Front. Psychol.* **7**, 902 (2016).

[6] Lyon, P. The biogenic approach to cognition. *Cogn. Process.* **7**, 11–29 (2006).

[7] Katz, Y. & Fontana, W. Probabilistic Inference with Polymerizing Biochemical Circuits. *Entropy* **24**, 629 (2022).

[8] Katz, Y., Springer, M. & Fontana, W. Embodying probabilistic inference in biochemical circuits. Preprint at https://doi.org/10.48550/arXiv.1806.10161 (2018).

[9] Cragg, B. G. & Temperley, H. N. Memory: the analogy with ferromagnetic hysteresis. *Brain J. Neurol.* **78**, 304–316 (1955).

[10] Keresztes, D. *et al.* Cancer drug resistance as learning of signaling networks. *Biomed. Pharmacother. Biomedecine Pharmacother.* **183**, 117880 (2025).

[11] Csermely, P. *et al.* Learning of Signaling Networks: Molecular Mechanisms. *Trends Biochem. Sci.* **45**, 284–294 (2020).

[12] Ginsburg, S. & Jablonka, E. Epigenetic learning in non-neural organisms. *J. Biosci.* **34**, 633–646 (2009).

[13] Kukushkin, N. V., Carney, R. E., Tabassum, T. & Carew, T. J. The massed-spaced learning effect in non-neural human cells. *Nat. Commun.* **15**, 9635 (2024).

[14] Doan, N., Theroux, A., Ramdas, T. & Gershman, S. J. Associative learning in the protozoan Stentor coeruleus.

[15] Rajan, D. *et al.* Single-cell analysis of habituation in Stentor coeruleus. *Curr. Biol. CB* **33**, 241-251.e4 (2023).

[16] Fields, C. & Levin, M. Competency in Navigating Arbitrary Spaces as an Invariant for Analyzing Cognition in Diverse Embodiments. *Entropy* **24**, 819 (2022).

[17] Pezzulo, G. & Levin, M. Top-down models in biology: explanation and control of complex living systems above the molecular level. *J. R. Soc. Interface* **13**, 20160555 (2016).

[18] Levin, M. The Computational Boundary of a "Self": Developmental Bioelectricity Drives Multicellularity and Scale-Free Cognition. *Front. Psychol.* **10**, 2688 (2019).

[19] Pashine, N., Hexner, D., Liu, A. J. & Nagel, S. R. Directed aging, memory, and nature's greed. *Sci. Adv.* **5**, eaax4215 (2019).

[20] Lahini, Y., Gottesman, O., Amir, A. & Rubinstein, S. M. Nonmonotonic Aging and Memory Retention in Disordered Mechanical Systems. *Phys. Rev. Lett.* **118**, 085501 (2017).

[21] Baulin, V. A. *et al.* Intelligent soft matter: towards embodied intelligence. *Soft Matter* **21**, 4129–4145 (2025).

[22] Loeffler, A. *et al.* Neuromorphic learning, working memory, and metaplasticity in nanowire networks. *Sci. Adv.* **9**, eadg3289 (2023).

[23] Kaspar, C., Ravoo, B. J., van der Wiel, W. G., Wegner, S. V. & Pernice, W. H. P. The rise of intelligent matter. *Nature* **594**, 345–355 (2021).

[24] Scott, H. L. *et al.* Evidence for long-term potentiation in phospholipid membranes. *Proc. Natl. Acad. Sci. U. S. A.* **119**, e2212195119 (2022).

[25] Abramson, C. I. & Levin, M. Behaviorist approaches to investigating memory and learning: A primer for synthetic biology and bioengineering. *Commun. Integr. Biol.* **14**, 230–247 (2021).

[26] Kouvaris, K., Clune, J., Kounios, L., Brede, M. & Watson, R. A. How evolution learns to generalise: Using the principles of learning theory to understand the evolution of developmental organisation. *PLoS Comput. Biol.* **13**, e1005358 (2017).

[27] Watson, R. A. & Szathmáry, E. How Can Evolution Learn? *Trends Ecol. Evol.* **31**, 147–157 (2016).

[28] Power, D. A. *et al.* What can ecosystems learn? Expanding evolutionary ecology with learning theory. *Biol. Direct* **10**, 69 (2015).

[29] Baldwin, J. M. A New Factor in Evolution. *Am. Nat.* **30**, 441–451 (1896).

[30] Hinton, G. E. & Nowlan, S. J. How Learning Can Guide Evolution. *Complex Syst.* **1**,.

[31] Smith, J. M. When learning guides evolution. *Nature* **329**, 761–762 (1987).

[32] Sridhar, V. H. *et al.* The geometry of decision-making in individuals and collectives. *Proc. Natl. Acad. Sci. U. S. A.* **118**, e2102157118 (2021).

[33] Strandburg-Peshkin, A., Farine, D. R., Couzin, I. D. & Crofoot, M. C. Shared decision-making drives collective movement in wild baboons. *Science* **348**, 1358–1361 (2015).

[34] Couzin, I. D. Collective cognition in animal groups. *Trends Cogn. Sci.* **13**, 36–43 (2009).

[35] Ward, A. J. W., Sumpter, D. J. T., Couzin, I. D., Hart, P. J. B. & Krause, J. Quorum decision-making facilitates information transfer in fish shoals. *Proc. Natl. Acad. Sci. U. S. A.* **105**, 6948–6953 (2008).

[36] Bazazi, S. *et al.* Collective motion and cannibalism in locust migratory bands. *Curr. Biol. CB* **18**, 735–739 (2008).

[37] Couzin, I. Collective minds. *Nature* **445**, 715 (2007).

[38] Couzin, I. D., Krause, J., James, R., Ruxton, G. D. & Franks, N. R. Collective memory and spatial sorting in animal groups. *J. Theor. Biol.* **218**, 1–11 (2002).

[39] Dreyer, T. *et al.* Comparing cooperative geometric puzzle solving in ants versus humans. *Proc. Natl. Acad. Sci.* **122**, e2414274121 (2025).

[40] Witkowski, O. & Ikegami, T. How to Make Swarms Open-Ended? Evolving Collective Intelligence Through a Constricted Exploration of Adjacent Possibles. *Artif. Life* **25**, 178–197 (2019).

[41] Piñero, J. & Solé, R. Statistical physics of liquid brains. *Philos. Trans. R. Soc. Lond. B. Biol. Sci.* **374**, 20180376 (2019).

[42] Turner, J. S. Termites as models of swarm cognition. *Swarm Intell.* **5**, 19–43 (2011).

[43] Sumpter, D. J. T. The principles of collective animal behaviour. *Philos. Trans. R. Soc. Lond. B. Biol. Sci.* **361**, 5–22 (2006).

[44] Lotka, A. J. Elements of Physical Biology. *Sci. Prog. Twent. Century 1919-1933* **21**, 341–343 (1926).

[45] Volterra, V. Fluctuations in the Abundance of a Species considered Mathematically1. *Nature* **118**, 558–560 (1926).

[46] Biswas, S., Clawson, W. & Levin, M. Learning in Transcriptional Network Models: Computational Discovery of Pathway-Level Memory and Effective Interventions. *Int. J. Mol. Sci.* **24**, 285 (2022).

[47] Biswas, S., Manicka, S., Hoel, E. & Levin, M. Gene regulatory networks exhibit several kinds of memory: Quantification of memory in biological and random transcriptional networks. *iScience* **24**, 102131 (2021).

[48] Pigozzi, F., Goldstein, A. & Levin, M. Associative conditioning in gene regulatory network models increases integrative causal emergence. *Commun. Biol.* **8**, 1027 (2025).

[49] Pla-Mauri, J. & Solé, R. Engineering Basal Cognition: Minimal Genetic Circuits for Habituation, Sensitization, and Massed–Spaced Learning. *ACS Synth. Biol.* **15**, 716–727.

[50] Virtanen, P. *et al.* SciPy 1.0: fundamental algorithms for scientific computing in Python. *Nat. Methods* **17**, 261–272 (2020).

[51] McInnes, L., Healy, J., Saul, N. & Großberger, L. UMAP: Uniform Manifold Approximation and Projection. *J. Open Source Softw.* **3**, 861 (2018).

[52] Pedregosa, F. *et al.* Scikit-learn: Machine Learning in Python. *J Mach Learn Res* **12**, 2825–2830 (2011).

[53] Ester, M., Kriegel, H.-P., Sander, J. & Xu, X. A density-based algorithm for discovering clusters in large spatial databases with noise. in *Proceedings of the Second International Conference on Knowledge Discovery and Data Mining* 226–231 (AAAI Press, Portland, Oregon, 1996).
[54] Elton, C. & Nicholson, M. The Ten-Year Cycle in Numbers of the Lynx in Canada. *J. Anim. Ecol.* **11**, 215 (1942).
[55] McElreath, R. *Statistical Rethinking: A Bayesian Course with Examples in R and STAN*. (Chapman and Hall/CRC, New York, 2020). doi:10.1201/9780429029608.
[56] Stein, R. R. *et al.* Ecological Modeling from Time-Series Inference: Insight into Dynamics and Stability of Intestinal Microbiota. *PLOS Comput. Biol.* **9**, e1003388 (2013)8. Katz, Y., Springer, M. & Fontana, W. Embodying probabilistic inference in biochemical circuits. Preprint at https://doi.org/10.48550/arXiv.1806.10161 (2018).
[57] Weinberg, S. Anthropic bound on the cosmological constant. Phys. Rev. Lett. 59, 2607–2610 (1987).
[58] Barrow, J. D. Cosmology, life, and the anthropic principle. Ann. N. Y. Acad. Sci. 950, 139–153 (2001).
[59] Silk, J. Teleological cosmology: the anthropic cosmological principle. Science 232, 1036–1037 (1986).
[60] Carter, B. Large Number Coincidences and the Anthropic Principle in Cosmology. Symp. - Int. Astron. Union 63, 291–298 (1974).
[61] De Houwer, J., Barnes-Holmes, D. & Moors, A. What is learning? On the nature and merits of a functional definition of learning. Psychon. Bull. Rev. 20, 631–642 (2013).
[62] Levin, M. Technological Approach to Mind Everywhere: An Experimentally-Grounded Framework for Understanding Diverse Bodies and Minds. Front. Syst. Neurosci. 16, 768201 (2022).
[63] McMillen, P. & Levin, M. Collective intelligence: A unifying concept for integrating biology across scales and substrates. Commun. Biol. 7, 378 (2024).